\documentclass[sigconf]{acmart} 
\AtBeginDocument{%
  }

\setcopyright{acmlicensed} 
\copyrightyear{2026} 
\acmYear{2026} 
\acmDOI{XXXXXXX.XXXXXXX} 
\acmConference[Conference acronym 'XX]{Make sure to enter the correct
  conference title from your rights confirmation email}{June 03--05,
  2018}{Woodstock, NY}  
\acmISBN{978-1-4503-XXXX-X/2026/08}  

\usepackage{tikz}
\usepackage{amsmath}
\usepackage{soul}
\usepackage{tcolorbox}
\tcbuselibrary{breakable,skins}
\usepackage{enumitem}
\usepackage{amsfonts}
\usepackage{xcolor}
\usepackage{algorithm} 
\usepackage{algpseudocode}

 \usepackage{multirow}
\usepackage{graphicx}
\usepackage{subcaption}
\usepackage{rotating}
\usepackage{adjustbox}
\usepackage{mwe}
\usepackage[htt]{hyphenat}
\usepackage{setspace}
\usepackage{float}
\usepackage{placeins}
\usepackage{listings}
\usepackage{booktabs}
\usepackage{caption}
\usepackage[table,xcdraw]{xcolor}
\usepackage{xurl}

\definecolor{BrickRed}{RGB}{203,65,84}
\definecolor{observationgreen}{RGB}{220, 255, 220}

\begin{document}

\title{Detokenization Leaks: \\Reconstructing Local LLM Outputs From Cache Traces}



\author{Roy Weiss}
\affiliation{%
\institution{Ben Gurion University}
  \country{Israel}}
\author{Benyamin Konstantinov}
\affiliation{%
  \institution{Ben Gurion University}
  \country{Israel}}
\author{Eitam Sheetrit}
\affiliation{%
  \institution{Microsoft Security}
  \country{Israel}}
\author{Tomer Simon}
\affiliation{%
  \institution{Microsoft Security}
  \country{Israel}}
\author{Yisroel Mirsky}
\authornote{Corresponding author}
\affiliation{%
  \institution{Ben Gurion University}
  \country{Israel}}
  
\renewcommand{\shortauthors}{Weiss et al.}

\begin{abstract}
We present a new attack that reconstructs the text generated by locally hosted LLMs by observing CPU cache activity during detokenization. Unlike prior attacks that rely on deployment-specific assumptions, such as shared data memory, CPU offloading, or Mixture-of-Experts architectures, our approach targets the detokenizer, a component used in default LLM inference pipelines. To obtain clean signals, we use Flush+Reload on shared tokenizer code to detect when decoding occurs, which lets us perform Prime+Probe at the right moment and isolate token-dependent cache activity. We then apply a clustering-and-language-model pipeline to recover text from noisy cache observations. We evaluate the attack across multiple datasets, hardware platforms, inference frameworks, and model families, and show that it can recover semantically accurate outputs from real-world local LLM deployments, including agentic systems. 

This vulnerability is particularly significant because the most widely used tokenizer implementations are susceptible to the attack and are embedded in many popular local LLM products and agent frameworks, including systems such as OpenClaw (which we demonstrate),
substantially broadening the practical attack surface.

\end{abstract}




\maketitle

\section{Introduction}
Large Language Models (LLMs) have rapidly transformed sectors such as healthcare, finance, education, and entertainment. Models such as ChatGPT and Llama provide high-quality natural language generation, summarization, and decision support across diverse applications. Increasingly, users rely on LLMs for sensitive tasks, including preliminary medical advice, drafting confidential communications, and financial planning \cite{raiaan2024review}. Beyond standalone usage, LLMs are now embedded in autonomous or semi-autonomous agent systems that orchestrate multi-step workflows, interact with external tools, and manage user data over extended sessions. Frameworks such as OpenClaw and Claude-Code enable agents to perform tasks including file management, coding, and personal assistance, often operating on sensitive inputs such as local files, credentials, and private communications \cite{wang2024survey, deng2025ai}.

\vspace{.3em}\noindent\textbf{A Shift to Local Hosting.}
As LLM adoption grows, there is a clear trend toward local deployment, driven by privacy, latency, and reduced reliance on external services. Advances in quantization, pruning, and optimized runtimes have made local execution feasible on consumer hardware \cite{ma2023llmpruner, kwon2023efficient, dettmers2023case}. Tools such as Llama.cpp \cite{llamacpp}, GPT4All \cite{gpt4all}, and Ollama \cite{ollama} allow users to run models including Llama \cite{grattafiori2024llama}, Mistral \cite{jiang2023mistral7b}, and Phi \cite{abdin2024phi} on desktops and laptops with minimal setup. This paradigm extends to agent-based systems, where users deploy LLM-powered agents locally to retain control over data and execution \cite{openclaw}. These agents typically run as long-lived processes with persistent access to local resources (e.g., files, APIs, and credentials), and therefore process sensitive information continuously over extended interactions.

\vspace{.3em}\noindent\textbf{Side-Channel Attacks on Local LLMs.} While local deployment improves privacy by keeping data on-device, it also shifts the threat model toward the underlying hardware. In particular, co-resident processes on the same machine may observe low-level microarchitectural side effects of LLM inference. One important class of such leakage arises from the CPU cache: because programs access memory through shared cache structures, an adversarial process can monitor cache contention to infer which memory locations another process is accessing. Since these memory access patterns depend on the data being processed, they can inadvertently reveal sensitive information. This creates an opportunity for adversaries to extract information directly from local LLM executions.

\vspace{.3em}\noindent\textbf{A Lack of Universality in the Threat Model.} Recent works have demonstrated cache-based side-channel attacks that exploit this effect to infer text generated by LLMs. However, these attacks rely on strong, deployment-specific assumptions. They require particular system or model features, such as unified CPU/GPU memory \cite{adiletta2025spill}, CPU layer offloading \cite{gao2025iknowwhatyousaid}, or the use of a Mixture-of-Experts (MoE) model architecture \cite{ding2025moecho}. Although these conditions can arise, some are rare, and none are \textit{universally} present across LLM deployments. For example, memory sharing of internal model data depends on specific system configurations, and many deployments use standard (non-MoE) architectures.

\vspace{.3em}\noindent\textbf{Proposed Approach.} In this paper, we propose a more broadly applicable cache-based side-channel attack, one that makes far fewer assumptions about the victim’s LLM libraries and configuration. Our attack is stronger because it targets a component that is \textit{universal} to all LLM pipeline deployments in their default configuration: the detokenizer. LLMs read and generate tokens, which are akin to words or word chunks. When a token is generated by the model, it is represented as an integer ID. To read the text, each ID must be mapped back to its corresponding word chunk. This occurs on the CPU via a lookup table, such as a hash map \cite{brown2020language, chowdhery2023palm}. Therefore, each time a token is generated, we expect to see a unique pattern in the cache that can be traced back to each token. 

\vspace{.3em}\noindent\textbf{Improved Universality.} To avoid relying on strong assumptions such as shared data memory, we instead employ a Prime+Probe attack to observe memory access patterns in the CPU cache. However, Prime+Probe is inherently noisy and is typically effective only when memory access patterns either repeat frequently or involve large memory regions that produce strong, distinguishable signals. In our setting, neither condition holds: each token’s decode operation touches only a small portion of the cache, resulting in weak and highly ambiguous signals that are difficult to isolate and interpret.

To address these challenges, we rely on two key insights that enable the effective use of Prime+Probe in this setting:

\begin{description}[leftmargin=1em]
\item[Insight 1.] \textit{Precise timing enables cleaner signals.} If we can align Prime+Probe measurements tightly around the detokenization step, we can significantly reduce noise and isolate the cache activity of interest. We achieve this using an auxiliary side channel: specifically, we apply Flush+Reload to the tokenizer library. In common deployments, tokenizer implementations (e.g., in \texttt{Llama.cpp}) are loaded as shared libraries by the operating system.\footnote{In contrast to \cite{adiletta2025spill, gao2025iknowwhatyousaid}, which assume shared data structures (often unavailable or disabled in standard deployments), we rely only on shared instruction pages (shared libraries), shared by default across processes in modern operating systems \cite{yarom2014flushreload}.} When the tokenizer code is invoked, it triggers a detectable signal, allowing us to precisely time our Prime+Probe measurements around each token decode operation.

\item[Insight 2.] \textit{Cache collisions are inevitable but structured.} Even with improved timing, cache signals remain noisy, and different tokens may map to overlapping cache sets, making them difficult to distinguish directly. To address this, we group tokens based on similarity in their observed cache access patterns and assign each group a discrete symbol. This produces a reduced symbolic representation of the trace sequence. Because these symbols preserve the order of generated tokens, we then train a downstream language model to map these symbol sequences to text, leveraging linguistic context to resolve ambiguities and reconstruct the original output.
\end{description}

\begin{figure}[t]
    \centering
    \includegraphics[width=\columnwidth]{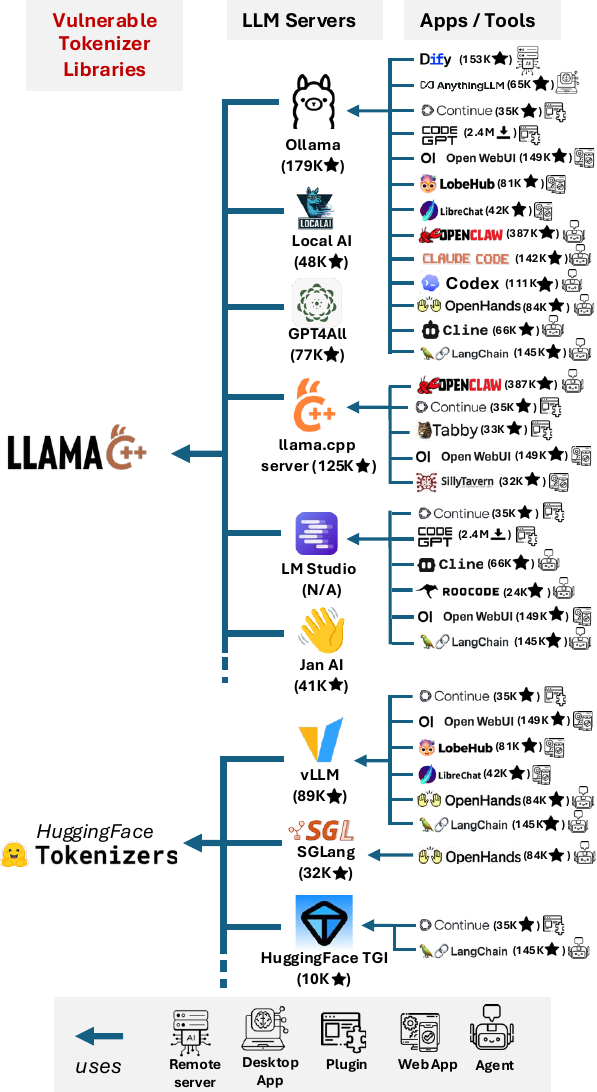}
    \caption{Tokenizers such as \texttt{Llama.cpp} and HuggingFace \texttt{Tokenizers} are vulnerable to our attack. Because these libraries are widely deployed in local LLM servers, the vulnerability extends to a large ecosystem of downstream applications, tools, and IDE plugins that depend on them (as of August 2026).}
    \label{fig:vuln_mapping}
\end{figure}

\vspace{.3em}\noindent\textbf{A Widespread Vulnerability.} Focusing on the tokenizer exposes a broad and practical attack surface. In current LLM ecosystems, a small number of tokenizer implementations (primarily \texttt{Llama.cpp} \cite{llamacpp} and HuggingFace Transformers \cite{wolf2020transformers}) are reused across a wide range of applications. As illustrated in Fig.~\ref{fig:vuln_mapping}, these libraries underpin numerous locally hosted systems, including LLM servers and agent frameworks, all of which rely on them as backend components.
Since these local instances run long-term, an adversary with minimal privileges can use this attack to (1) profile a target instance, (2) learn the mapping between cache traces and tokens using our method, and (3) subsequently reconstruct future outputs from that instance. This enables the extraction of sensitive user information across a diverse set of real-world deployments.

In our evaluation, we show that the attack can decipher content from a wide range of LLM tasks, from conversational LLMs to instruct LLMs to code LLMs, using three datasets (UltraChat~\cite{ding2023ultrachat}, ChatDoctor~\cite{li2023chatdoctor}, and Code-Alpaca~\cite{codealpaca}). We also evaluate our attack across two hardware platforms (laptop and desktop), two inference frameworks, and two target models, covering diverse tokenizer implementations and application domains. We also evaluate an end-to-end attack on the popular locally hosted agent OpenClaw~\cite{openclaw}.

\vspace{.5em}\noindent\textbf{Contributions.} In summary, this paper makes the following contributions:

\begin{itemize}[leftmargin=1em]
    \item \textbf{A More Universal Cache-Based Side-Channel Attack.} We present a detokenization-based side-channel attack that operates under default system configurations, avoiding the strong assumptions required by prior work (e.g., shared data structures, specific model architectures, or specialized CPU/GPU memory configurations). This significantly broadens the applicability of cache-based attacks on local LLM deployments.
    
    \item \textbf{Precise Prime+Probe via Auxiliary Synchronization.} We show how to obtain fine-grained temporal alignment for Prime\allowbreak+\allowbreak Probe by leveraging Flush+Reload on shared tokenizer code. This enables targeted measurement of detokenization events and substantially improves signal quality despite the inherent noise of Prime+Probe.
    
    \item \textbf{Robust Trace-to-Text Reconstruction from Noisy Signals.} We introduce a novel pipeline for mapping noisy, collision-prone cache traces to natural language. Our approach clusters continuous timing data into symbolic sequences and uses language models to reconstruct text, leveraging context to resolve ambiguity.
    
    \item \textbf{Comprehensive Evaluation on Real-World Systems.} We evaluate our attack across diverse LLM tasks, hardware platforms, tokenizer frameworks, and model families, and demonstrate end-to-end attacks on real-world deployments, including a popular locally hosted agent system (OpenClaw).
\end{itemize}

\section{Background}
In this section, we review the LLM inference pipeline and the cache side-channel primitives relevant to our setting, focusing on decode-time behavior and shared CPU resources. These components together define the attack surface exploited in our setting.

\subsection{LLM Inference and Tokenization \label{sec:background_llm_inferece}}
LLMs operate on discrete units of text known as \emph{tokens}, which typically correspond to words, subwords, or characters. During inference, the model consumes a user-provided input prompt and generates a sequence of output tokens in an autoregressive manner. This process involves two distinct stages: the \textit{encode} phase and the \textit{decode} phase.

In the \textit{encode} phase, the input prompt string \( p \in \Sigma^* \) is tokenized on the CPU into a sequence of token IDs \((p_1, p_2, \ldots, p_n) \in V^n \), using a deterministic vocabulary (\(V\)) defined by the tokenizer. These IDs are then passed to the model as the initial inference context.

The \textit{decode} phase maps each token ID to its corresponding string. At each step, the model predicts the next token ID \( r_i \) based on the previously generated context. To present the output in human-readable form, these integer token IDs are converted to UTF-8 strings using the \textit{decode table} (\(D\)), which resides on the CPU, in contrast to the model parameters that typically remain on the GPU. To provide real-time feedback, inference frameworks invoke the decode operation immediately after each token is generated, making this mapping process inherently autoregressive.

\subsection{Local LLM Deployments}

Advances in quantization, pruning, and optimized inference runtimes have enabled LLMs to run directly on commodity personal devices \cite{ma2023llmpruner, kwon2023efficient, dettmers2023case}. Motivated by latency and privacy, local deployments are increasingly used in developer tools, document processing systems, and personal assistants \cite{xu2025fast, yuan2024wip, zhang2024enabling}.

Most local deployments follow a client-server architecture in which an inference backend exposes a localhost API to desktop interfaces, IDE plugins, or automation tools. Clients submit prompts and receive responses, typically streamed token by token as they are generated and decoded. To avoid repeatedly loading model weights, the backend remains active and serves multiple requests over time. Systems such as \texttt{Ollama} and \texttt{LM Studio} follow this design \cite{ollama}.

\subsection{Cache Side-Channel Attacks \label{sec:background_cache}}
Modern CPUs employ caches to reduce memory access latency. These caches are organized into multiple levels, typically L1, L2, and the Last-Level Cache (LLC). L1 and L2 caches are private to individual cores, while the LLC is shared across cores. On each memory access, the CPU first checks the cache hierarchy: a cache \textit{hit} returns data quickly, while a \textit{miss} fetches it from main memory and stores it in the cache. This shared caching behavior creates observable contention between processes, enabling an attacker to infer a victim’s memory access patterns from a separate process.

\vspace{0.3em}\noindent\textbf{Flush-based Attacks.}
Flush-based side-channel techniques, such as Flush+Reload \cite{yarom2014flushreload} and Flush+Flush \cite{gruss2016flushflush}, exploit scenarios where memory pages are shared between an attacker and a victim. In Flush+Reload, the attacker evicts a shared cache line using \texttt{clflush} and measures the time to reload it; a fast access indicates that the victim accessed the same line.

This enables high-resolution monitoring of specific cache lines corresponding to code or data. In practice, shared memory arises from shared libraries (e.g., dynamically linked code). Extending this to arbitrary data structures requires additional conditions, such as explicit shared mappings, page deduplication, or unified memory. As these are not consistently present, the applicability of Flush+Reload-style attacks on data is limited.

\vspace{0.3em}\noindent\textbf{Prime+Probe Attacks.}
Prime+Probe is a contention-based side-channel technique that does not require any shared memory between attacker and victim. Instead, it exploits the fact that multiple addresses may map to the same cache set, creating measurable interference. The attacker first fills selected cache sets with their own data (prime phase), allows the victim to execute, and then measures access times to the same addresses (probe phase). Increased access latency indicates that the victim has evicted some of the attacker’s cache lines, revealing information about the victim’s memory accesses.

\vspace{0.3em}\noindent\textbf{L1 Prime+Probe.}
In our setting, we focus on the L1 data cache, which is private to a physical core but shared between simultaneous multithreading (SMT) siblings. This enables an attacker co-scheduled on the same physical core to monitor L1 cache activity with high temporal fidelity and low measurement latency, making it well-suited for short per-token decode operations in autoregressive LLM inference. On common x86 systems, L1 set selection is determined by the virtual address bits within the page offset, so accesses to different entries may contend on the same set even without shared memory. At the same time, L1 Prime+Probe is challenging: the cache is small, the hit-miss timing gap is narrow, and the number of sets is limited. As a result, measurements are noisy and multiple memory locations may alias to the same observed cache set, reducing spatial resolution \cite{rauscher2025systematic}.

\subsection{Decode Tables in Practice \label{sec:background_decode_tables}}
The \textit{decode table} \(D\) is a data structure that maps token identifiers to their corresponding UTF-8 byte sequences. In mainstream implementations, it is stored as an array-like structure indexed by token ID. For example, in \texttt{Llama.cpp}, which underlies frameworks such as \texttt{Ollama} and \texttt{GPT4All}, it is implemented as a \texttt{std::vector}. Other libraries use equivalent containers. In HuggingFace's \texttt{Tokenizers} library, for example, the decode table is implemented using Rust's default \texttt{HashMap}. Although its randomized hash seed may cause the memory layout to differ across process invocations, the layout remains fixed once initialized. Thus, within a long-lived process, decoding the same token consistently accesses the same memory locations, producing stable and repeatable cache access patterns.

\section{Attack Overview \label{sec:attack_overview}}

\subsection{Threat Model \label{sec:threat_model}}

\vspace{0.3em}\noindent\textbf{Victim Setup.}
We consider a user running an LLM locally on a personal device, either as a standalone server or as the backend of a local application or agent framework. Such deployments are increasingly used for sensitive tasks, including summarizing emails, drafting confidential communications, providing personal advice, and managing files and services.

This setting appears in several common forms: (i) local model servers used by developer tools, (ii) desktop assistants that optionally expose a localhost API, and (iii) workstation inference services shared by multiple local clients. In practice, these deployments typically run a long-lived backend process that exposes an API accessible over \texttt{localhost}. For example, \texttt{Ollama} binds to \texttt{127.0.0.1:11434} by default and exposes an HTTP API such as \texttt{POST /api/generate}, through which any local process can submit prompts and receive responses. Higher-level applications and agent frameworks connect to such local endpoints. For example, \texttt{OpenClaw} supports local model providers including \texttt{Ollama} and other local inference backends.

\textit{Persistence and execution.} We assume that inference runs locally on the host CPU and/or GPU and that the LLM instance is long-lived, serving multiple requests over time. This ensures that tokenizer data structures remain stable during execution, as described in Section~\ref{sec:background_decode_tables}.

\textit{Platform assumptions.} We assume the victim system employs simultaneous multithreading (SMT), such as Intel Hyper-Threading, exposing multiple logical processors per physical core. This is a practical assumption, as SMT is commonly enabled by default on consumer systems \cite{intel_ht_gaming}, including across a broad range of modern CPUs summarized in Appendix Table~\ref{appendix:cpu-table}. Moreover, disabling SMT can reduce performance by 25\%--35\% \cite{walton2019intel}, providing a practical incentive to leave it enabled. Under SMT, the adversary can identify the processor executing the victim and place its own thread on the corresponding sibling processor using standard unprivileged affinity mechanisms, as commonly assumed in SMT-based side-channel attacks \cite{bhattacharyya2019smotherspectre, gast2023squip}.

Finally, we assume that the tokenizer implementation includes executable code pages that are shared across processes, as is standard for dynamically linked shared libraries and native extension modules in modern operating systems. We verified this property for the popular libraries we evaluated, including HuggingFace's \texttt{Transformers} library and \texttt{Llama.cpp} as shown in Figure \ref{fig:vuln_mapping}. Our attack does not assume that tokenizer data structures are shared; it relies only on shared code pages, which are enabled by default.

\vspace{0.3em}\noindent\textbf{Adversary Model.}
The adversary’s objective is to reconstruct the plaintext output generated by the hosted LLM instance, or more generally, to infer sensitive information from that output.
To achieve this objective, we consider an unprivileged adversarial process co-located on the same machine as the victim and scheduled on the corresponding SMT sibling. The adversary runs as a benign user-space application (e.g., a desktop app or IDE plugin) and does not require elevated privileges or modifications to the operating system or LLM framework.

\textit{Capabilities.} The adversary can (1) execute standard user-space instructions, including cache line flush (\texttt{clflush}) and high-resolution timing (\texttt{rdtsc}), (2) access shared code pages of dynamically linked libraries used by the victim, and (3) interact with the local LLM service by issuing prompts and observing outputs through its exposed local interface. None of these capabilities requires special privileges. As is standard in profiling-based side-channel attacks, the adversary can use this access to collect labeled traces and learn a mapping between cache observations and generated text before applying the attack to unknown executions \cite{ding2025moecho, zhang2024r+, yuan2022automated, yuan2021private}.

\textit{Knowledge.} Our attack is agnostic to the victim's model family, model version, quantization, vocabulary size, inference framework, and application configuration. The only implementation-specific information required is the tokenizer library used by the victim, since its shared decode routine provides the Flush+Reload trigger. This library need not be known a priori: as described in Section \ref{sec:attack_implementation}, the attacker can identify the active tokenizer implementation by monitoring candidate decode routines while inducing generation. These assumptions are consistent with prior cache side-channel threat models \cite{yarom2014flushreload, liu2015llc}.

\textit{Exclusions.} The adversary does not require access to shared data structures, internal model states, or GPU memory. Unlike prior work~\cite{adiletta2025spill, gao2025iknowwhatyousaid}, we do not assume any shared data memory between attacker and victim. The attack relies only on shared instruction pages from dynamically linked libraries, which are read-only and shared by default in modern operating systems.

\subsection{Vulnerability Breakdown \label{sec:vuln}}

\noindent\textbf{Key insight.}
The core vulnerability is that LLM detokenization relies on a \emph{deterministic decode table} whose memory access pattern depends on the generated token. During inference, each output token \(r_i\) is represented internally as an integer ID, and the tokenizer’s decode function maps that ID to its corresponding UTF-8 string by reading the decode table \(D\). Because different tokens correspond to different entries in \(D\), decoding different tokens causes accesses to different memory locations. As discussed in Section~\ref{sec:background_decode_tables}, this mapping remains fixed for the lifetime of the process. As a result, the same token repeatedly induces the same cache activity, creating a stable leakage source that can be profiled and later exploited.

\vspace{0.3em}\noindent\textbf{Why this matters.}
This leakage arises from a component that is fundamental to LLM inference: every generated token must eventually be converted back into text. Unlike prior attacks that depend on deployment-specific features such as shared data memory or particular model architectures, our leakage source comes from the decode-table lookup itself, making the attack applicable across a broad range of local LLM deployments.

\vspace{0.3em}\noindent\textbf{Technical formulation.}
As discussed in Section~\ref{sec:background_llm_inferece}, LLMs operate over a fixed vocabulary of token identifiers. During generation, each predicted token \(r_i\) must be decoded into a human-readable string by querying the decode table \(D\). Since different token strings reside at different memory locations, the memory access pattern of the decode operation depends on \(r_i\).

Although one process cannot directly observe another process's memory accesses, each lookup of \(r_i\) induces activity in an L1 cache set \(c_j\) that can be detected via Prime+Probe. The mapping from a token identifier \(r_i\) to its cache set is given by
\[
c_j = \left( \left((r_i \cdot E) \bmod P \right) \gg \log_2 B \right) \,\&\, (N - 1),
\]
where \(E\) is the size of an entry in the decode table \(D\), \(P\) is the page size, \(B\) is the cache line size, and \(N\) is the number of L1 cache sets. On typical x86 systems (\(B=64\), \(N=64\)), this simplifies to
\[
c_j = \left( (r_i \cdot E \bmod P) \gg 6 \right) \,\&\, 63.
\]

As detailed in Section~\ref{sec:background_decode_tables}, the decode-table layout remains fixed after initialization, so the mapping from a token identifier \(r_i\) to its corresponding cache set \(c_j\) remains stable during execution. This enables a profiling-style attack: the adversary collects a limited set of \emph{labeled} traces from the same long-lived victim by issuing benign queries and observing the corresponding outputs, producing trace--token pairs that capture decode-table accesses. These pairs are then used to learn a mapping from cache traces to tokens and reconstruct outputs generated for other applications or users of the same server.

Let \(t_i \in \mathbb{N}^N\) denote the cache trace collected during the decode step of token \(r_i\), where each component records the access latency of one L1 cache set. The trace \(t_i\) exhibits a token-dependent signature, with its dominant signal at index \(c_j\). Observing the sequence \((t_1, t_2, \ldots)\) therefore reveals information about the ordered token sequence \((r_1, r_2, \ldots)\), enabling reconstruction of future outputs from traces alone.

\begin{tcolorbox}[colback=observationgreen,colframe=green!80!black,fonttitle=\bfseries,boxsep=1mm, bottom=0mm, left=1mm, right=1mm, top=0mm]
\textbf{Insight.} Detokenization exposes an output-aligned side channel. Each generated token is decoded in output order, so the resulting cache observations form a noisy representation of the generated token sequence itself.
\end{tcolorbox}


However, exploiting this leakage is non-trivial due to several inherent challenges associated with Prime+Probe:

\vspace{0.3em}\noindent\textbf{Challenge 1 (C1):} \textit{Cache Collisions.}  
The L1 cache of a consumer CPU has only about 64 cache sets that an attacker can monitor with Prime+Probe. Yet each set may correspond to an enormous number of memory addresses. For example, on a system with 8GB of RAM, over 100 million addresses can map to the same set. This massive collision space makes it extremely difficult to isolate and reliably monitor a single address. As a result, naively applying Prime+Probe to monitor $D$ is impractical.

\vspace{0.3em}\noindent\textbf{Challenge 2 (C2):} \textit{Low signal-to-noise ratio.} 
In the L1 cache, the timing difference between a hit and a miss can be only a few cycles, making measurements highly sensitive to noise. OS activity and scheduling interference further reduce signal clarity. Since our attacker operates in a single-shot setting (the generated token keeps changing), averaging across repeated measurements (as done in other works) is not possible.

\vspace{0.3em}\noindent\textbf{Challenge 3 (C3):} \textit{Coarse spatial resolution.} 
The number of L1 cache sets \(N\) is small relative to the vocabulary size \(|V|\). Because Prime+Probe observes activity only at cache-set granularity, the mapping from tokens to sets is many-to-one. Multiple tokens may therefore map to the same cache set, creating aliasing and ambiguity in reconstruction.

To make Prime+Probe effective despite these limitations, we address \textbf{C1} by performing Prime+Probe at the moment a token is about to be decoded. By narrowing the probe window, we eliminate a significant number of collisions that occur due to other processes. We accomplish this with an auxiliary side channel based on \texttt{Flush+Reload}, which is used to detect when the victim invokes the decode function, upon which the spy process triggers a Prime+Probe measurement to capture the resulting cache activity (§\ref{sec:detection_impl}). To address \textbf{C2}, we characterize the timing behavior of the decode routine and use it to choose when probing begins and the delays between Prime+Probe steps. This improves overlap with the victim's token-dependent accesses and yields more stable measurements (§\ref{sec:primeprobe_impl}). Finally, to mitigate the ambiguity introduced by \textbf{C3}, we employ a language model-based reconstruction method that translates noisy cache traces into semantically meaningful token sequences (§\ref{sec:llm_impl}).

\subsection{Attack Workflow \label{sec:attack_workflow}}

At a high level, the attack has four steps: (1) wait for the LLM to be prompted, (2) detect when the next generated token is about to be decoded using Flush+Reload (Section \ref{sec:detection_impl}), (3) collect cache traces for the current token using Prime+Probe (section \ref{sec:primeprobe_impl}, repeat steps 2-3 for the entire response, (4) translate traces into text using a predictive model (Section \ref{sec:llm_impl}). 

\begin{figure}[t]
    \centering
    \includegraphics[width=\columnwidth]{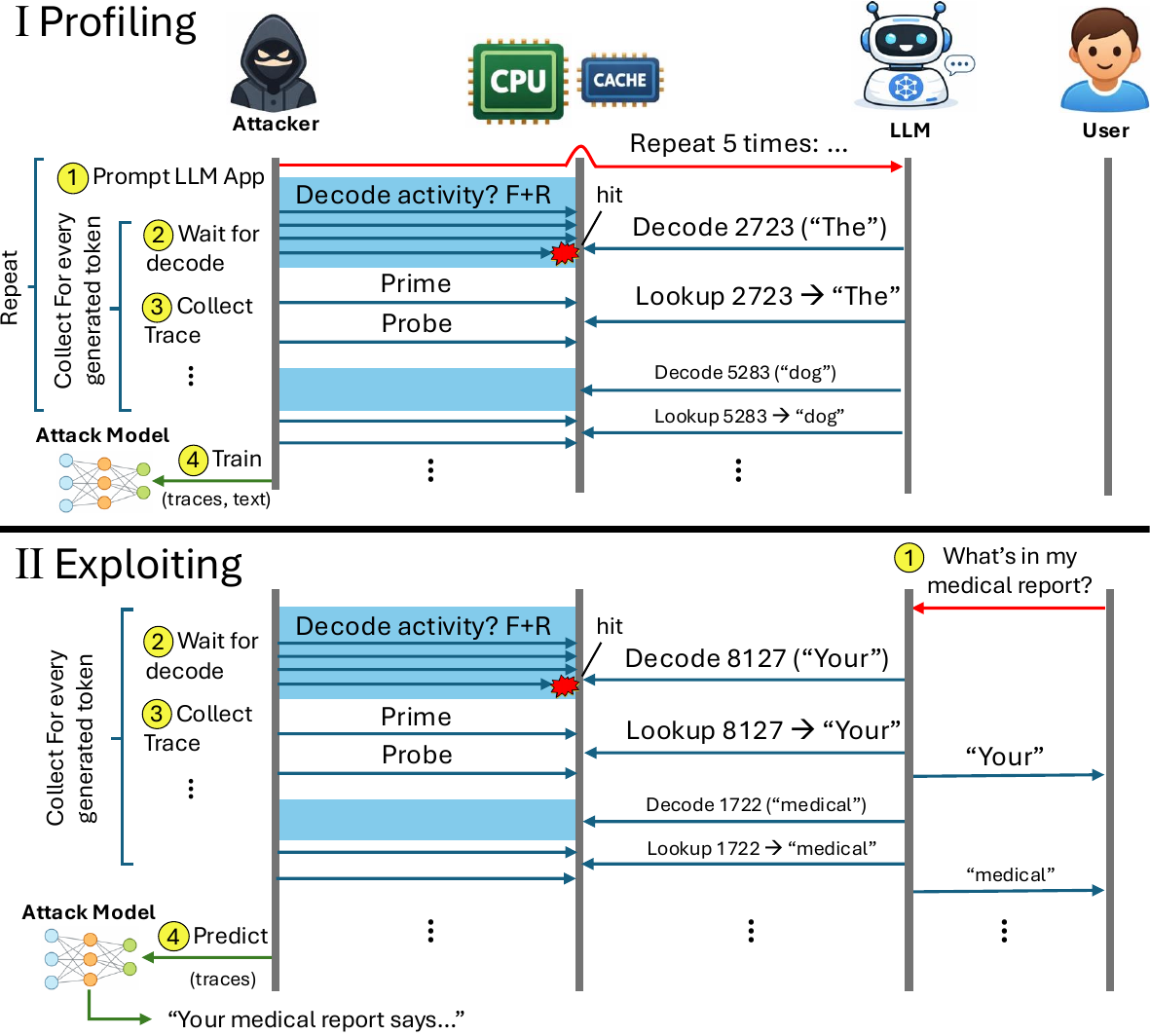}
    \caption{Attack workflow illustrating the profiling and exploitation phases and their 4 steps. During profiling, the adversary induces generation to collect traces covering common tokens and trains a predictive model; during exploitation, it uses this model to reconstruct future conversations.}
    \label{fig:attack_overview}
\end{figure}

To realize this attack, the adversary must first obtain the predictive model used for reconstruction. We therefore divide the attack into two phases, illustrated in Figure~\ref{fig:attack_overview}: a \emph{profiling} phase, in which the adversary learns the mapping from cache traces to text for a specific long-lived server instance, and an \emph{exploitation} phase, in which the learned model is used to reconstruct future outputs from the same instance.

\vspace{0.3em}\noindent\textbf{Phase I: Profiling.}
The adversary first interacts with the local LLM service through its exposed API and issues its own prompts, causing the model to generate tokens whose ground-truth text is known to the adversary. In parallel, the adversary monitors decode activity using \texttt{Flush+Reload} and collects a cache trace using Prime+Probe \textit{for each} generated token. This produces a labeled dataset that maps generated tokens to their corresponding cache traces. The adversary then trains a reconstruction model on this dataset. As shown in Section~\ref{sec:query_tradeoff}, profiling can be performed rapidly using only 250 targeted queries, or using approximately 7,650 ordinary conversational queries to obtain an equivalent profile when less distinctive interactions are desired.

\vspace{0.3em}\noindent\textbf{Phase II: Exploitation.}
After profiling, the adversary passively monitors the same server instance while it serves other users or applications. The adversary again uses \texttt{Flush+Reload} to detect decode invocations and, on each detection, performs a Prime+Probe measurement to capture token-level cache activity. Repeating this over the full response yields a sequence of cache traces, which is then fed to the trained reconstruction model to recover the generated text.

In the next section, we will carefully detail how we designed and implemented Steps 2–4.

\section{Attack Implementation \label{sec:attack_implementation}}

\subsection{LLM Prompting (step 1)}
During exploitation (Phase II), the attacker simply proceeds to Step 2 and waits for a user or application to prompt the target LLM server. During profiling (Phase I), the attacker instead actively queries the server to obtain traces paired with ground-truth response tokens. We consider two profiling approaches. In \textit{targeted profiling}, the attacker asks the model to repeat selected token sequences, providing controlled token coverage and multiple traces per token with few queries. For example, we use queries of the form \textit{``repeat the following sequence 5 times: $\langle$sequence of 10 tokens$\rangle$''}. Alternatively, in \textit{benign profiling}, the attacker issues ordinary conversational queries and collects traces from the tokens that naturally occur in the responses. As shown in Section~\ref{sec:query_tradeoff}, targeted profiling achieves strong reconstruction with only 250 queries, while benign profiling reaches comparable performance using a larger number of ordinary interactions.

\subsection{Detecting an Imminent Token (step 2) \label{sec:detection_impl}}
In this step, we determine when a token is about to be decoded using $D$ in order to probe the cache at the right moment. We do so by detecting when the decode routine is executed. Building on prior work~\cite{yarom2014flushreload}, we use an auxiliary side channel based on the Flush+Reload technique over shared libraries. Since tokenizer implementations are loaded as shared libraries, their instruction pages are shared across processes, enabling reliable detection of decode execution across all evaluated frameworks (Figure~\ref{fig:vuln_mapping}).

To set up this `trigger' side channel, the adversary first identifies the tokenizer library used by the victim and the address of its decode routine. This does not require knowledge of the victim's model or inference configuration. Common tokenizer implementations are typically open-source or publicly distributed and loaded through shared executable pages, as is standard on modern operating systems. The adversary can therefore monitor candidate decode routines while issuing prompts to the local service; the routine whose Flush+Reload activity correlates with generated output reveals the active tokenizer implementation. Once identified, the routine can be analyzed offline using standard tools such as \texttt{gdb} or \texttt{readelf}. For example, in \texttt{Llama.cpp}, the adversary may locate the symbol \texttt{llama\_detokenize} at a fixed offset within the shared object file.\footnote{This holds for version 0.3.16 of the \texttt{Llama.cpp} Python binding library, available at \url{https://github.com/abetlen/llama-cpp-python}} The adversary then maps the shared library into its own address space (e.g., via \texttt{mmap} or \texttt{dlopen}) to obtain the base virtual address, computes the target address, and monitors it using Flush+Reload. Concretely, the attacker repeatedly flushes the target cache line using \texttt{clflush} and measures its reload time; when the victim executes the decode function, the line is brought back into the cache, resulting in a significantly lower access latency that serves as a trigger signal. This process is unaffected by ASLR since the relevant code remains \emph{physically shared} across processes.

We implement the detection phase using a widely adopted side-channel toolkit~\cite{yarom2016mastik}. The target address is monitored in a tight loop with minimal pauses to ensure high temporal resolution. Since Flush+Reload operates at cache-line granularity, the signal remains localized, improving the signal-to-noise ratio~\cite{yarom2014flushreload, rauscher2025systematic}.

\noindent\textbf{Experiment.} To evaluate the detection performance, we simulate decode invocations at random intervals of 10--30\,ms, reflecting typical token-generation delays. Over 1.2 million invocations, an attacker process continuously monitors the decode function and achieves a 97.66\% true positive rate with zero false positives, demonstrating high reliability.

\subsection{Capturing a Token's Cache Trace (step 3)\label{sec:primeprobe_impl}}

If a decode event is detected from step 2, a token is \textit{about} to be decoded. At this point, the attacker captures a trace of the decode-table lookup in the cache. Since the decode table itself is not shared between processes, Flush+Reload cannot be used to monitor its token-dependent accesses. We therefore use Prime+Probe.

When the decode of token $r_i$ is detected, we use Prime+Probe to monitor all $N = 64$ sets of the L1 cache, obtaining one trace ($t_i \in \mathbb{N}^{64}$) of access latencies. 
Although the attacker can reliably detect the invocation of the decode function, the precise timing of the table lookup within \(D\) remains unobservable. Therefore, through an offline analysis, the adversary optimizes the Prime+Probe measurement using high-resolution \texttt{rdtsc} timing. 
In our desktop \texttt{Llama.cpp} setup, for example, we found that the ideal time window is to probe from \(8\) $\mu$s after the trigger event for a total of \(15\) $\mu$s. This delay remains stable across models within the same framework, such as Phi and Llama-3.

We also experimented with applying Prime+Probe to the LLC, but found it impractical for this leakage source. Unlike the 64-set L1 cache, the LLC is orders of magnitude larger and substantially slower to probe \cite{liu2015llc, rauscher2025systematic}. Moreover, the decode table spans a large memory region whose entries are distributed broadly across the LLC, requiring the attacker to monitor a large number of cache sets. Covering this set space within the short decode window was therefore not feasible.

\begin{figure}[t]
  \centering
  \includegraphics[width=\linewidth]{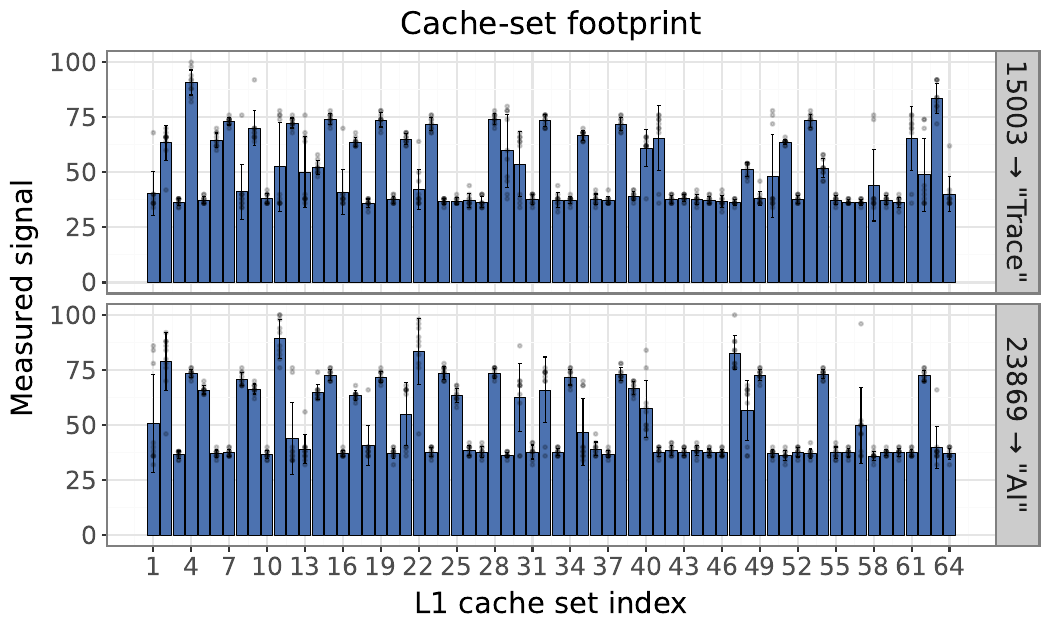}
  \caption{Example cache traces collected during the \texttt{decode} operation of two tokens. Each bar shows the average access latency of a cache set over 10 repetitions.}
  \label{fig:token_cache_trace}
\end{figure}

Figure~\ref{fig:token_cache_trace} shows representative single probe traces for two tokens, illustrating how different decode operations induce distinct cache-set traces.

We briefly note several implementation details that are important for obtaining stable measurements. We construct eviction sets covering all L1 cache sets using virtual addresses that map to distinct indices, with each set represented by a small group of congruent addresses. L1 Prime+Probe measurements are inherently noisy due to hardware prefetching, limited cache capacity, and self-interference. We found that existing toolkits such as Mastik \cite{yarom2016mastik} do not provide sufficiently stable measurements in this regime. Instead, we use a modified version of \texttt{CacheSC} \cite{haller2020revisiting}, which offers finer control over memory access patterns. In this implementation, eviction sets are traversed as randomized linked lists to avoid predictable access patterns and reduce both hardware prefetching effects and self-evictions, significantly improving signal quality.

Overall, this procedure yields one trace \(t_i\) per decoded token, forming the trace sequence \((t_1, t_2, \dots)\) for the entire generated response.

\begin{figure*}[t]
  \centering
  \includegraphics[width=\textwidth]{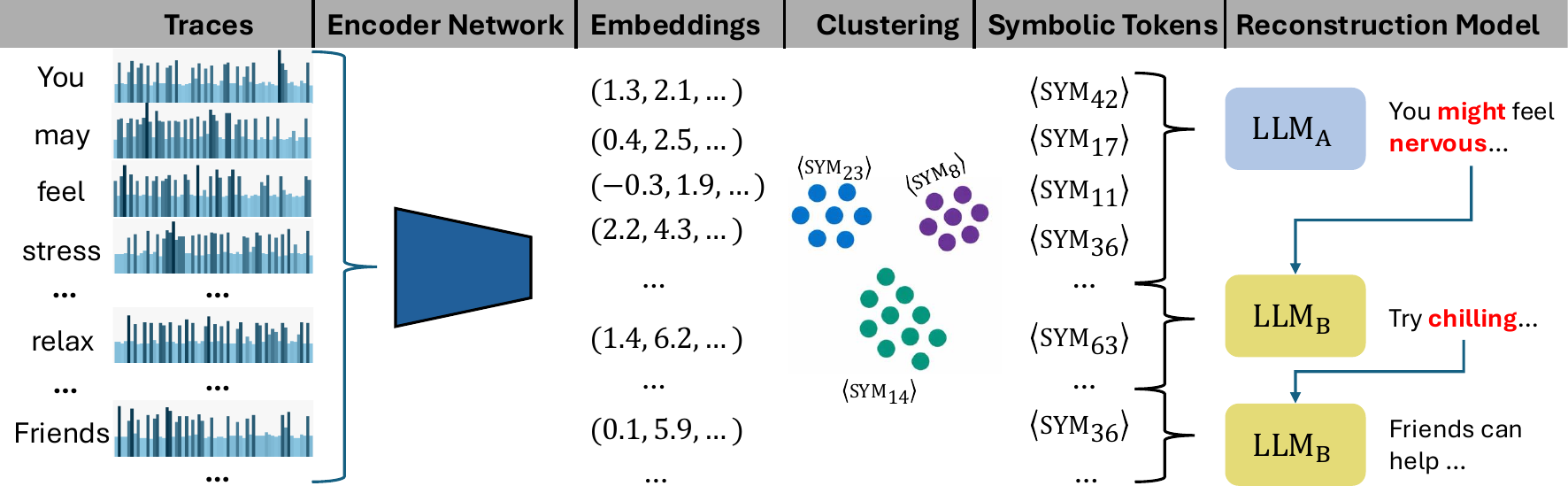}
  \caption{The trace-to-text reconstruction pipeline. Each trace in the sequence is first encoded into an embedding and assigned to a cluster learned during the profiling phase. The trace is then replaced with the symbol corresponding to that cluster, producing a sequence of symbols. Finally, this symbol sequence is passed to a language model trained to reconstruct the original tokens. Following prior work, we use separate models for the first segment (LLM$_A$) and subsequent segments (LLM$_B$).}
  \label{fig:reconstruction_pipeline}
\end{figure*}

\subsection{Text Inference via Predictive Model (step 4) \label{sec:llm_impl}}

After collecting traces $(t_1, t_2, \dots)$, the adversary’s final step is to reconstruct the original response tokens $(r_1, r_2, \dots)$. This is challenging because thousands of tokens are compressed into just 64 observable cache sets, leading to severe collisions that make direct trace-to-token recovery infeasible (\textbf{C3}). The problem is further compounded by timing noise and limited measurement resolution (\textbf{C1, C2}). Crucially, however, these collisions are not random: they are stable and structured. Using clustering, we map groups of tokens with similar trace patterns to shared symbols. This converts the sequence of numerical traces into a symbolic sequence that serves as a coarse representation of $(r_1, r_2, \dots)$. Although lossy, this symbolic sequence preserves substantial linguistic structure, which is sufficient for us to train a language model to reconstruct the original text. An overview of our reconstruction pipeline is shown in Figure~\ref{fig:reconstruction_pipeline}.

\begin{tcolorbox}[colback=observationgreen,colframe=green!80!black,fonttitle=\bfseries,boxsep=1mm, bottom=0mm, left=1mm, right=1mm, top=0mm]
\textbf{Insight.} Although individual Prime+Probe traces are ambiguous because of heavy collisions, this ambiguity is reduced when traces are viewed as a sequence and constrained by the fact that they must correspond to coherent language. 
\end{tcolorbox}

We now describe how this pipeline is instantiated in practice, focusing on (i) mapping traces to symbolic tokens, and (ii) reconstructing text from the resulting symbol sequences.

\subsubsection{\textbf{Trace Symbolization}}\label{sec:symbolization}
As a preliminary step, we evaluate whether individual traces can be mapped directly to tokens. We train a multilayer perceptron (MLP) classifier to predict token identifiers from raw latency vectors. The classifier achieves very low accuracy (e.g., below 1\% top-1), indicating that individual traces provide weak token-level information due to cache-set aliasing and measurement noise. This motivates the use of an intermediate representation that groups traces with a similar structure.

Each trace \(t_i\) is represented as a 64-dimensional latency vector, one value per monitored L1 cache set. We normalize these vectors and map them to a normalized embedding space using a lightweight MLP encoder trained with a supervised-contrastive objective~\cite{khosla2020supervised}, where traces of the same token are treated as positives. We then compute one centroid per token by averaging its profiling-trace embeddings and cluster the token centroids into \(K\) groups using K-means with cosine similarity~\cite{lloyd1982least}.

We learn an embedding before clustering rather than applying K-means directly to raw traces, which are noisy and overlap substantially in the original latency space. Using token-level supervision available during profiling, the encoder learns to group traces of the same token and separate those of different tokens, improving cluster separation and downstream symbolization.

Each cluster defines a symbolic token \(s_\ell \in S\), where \(S=\{s_1,\dots,s_K\}\). During symbolization, each trace is assigned to its nearest cluster center and replaced by the corresponding symbol, producing a symbolic sequence that captures the coarse structure of the underlying token sequence. We use \(K=64\), which balances clustering fidelity and symbol diversity (Appendix~\ref{app:k-ablation}).

To integrate these symbols into the reconstruction model, we extend the tokenizer vocabulary with \(K\) synthetic entries of the form \(\langle \mathrm{SYM}_k \rangle\), one per cluster, following prior work~\cite{weiss2024your}. Each is assigned a unique token ID appended to the base vocabulary, allowing symbolic sequences to be processed through the model's standard embedding layer.

\subsubsection{\textbf{Semantic Reconstruction}}

Given a symbolized trace sequence \((s_1, s_2, \dots)\), we reconstruct the corresponding response tokens \((r_1, r_2, \dots) \in V^*\) by treating the task as sequence-to-sequence translation from symbolic tokens to natural language using Flan-T5-XL~\cite{flant5}.

We follow Weiss et al. \cite{weiss2024your}, who observe that the first sentence of LLM responses often reveals the main topic and facilitates reconstruction of subsequent content. Accordingly, they employ two separate models for staged reconstruction, a design we adopt. Our contribution instead lies in extracting informative discrete symbol sequences from noisy microarchitectural traces.

We split each response into 32-token segments \(R=(R_0, R_1, \dots)\). \(\mathrm{LLM}_A\) reconstructs the first segment \(R_0\) from its symbol sequence alone. For \(j>0\), \(\mathrm{LLM}_B\) reconstructs segment \(R_j\) conditioned on the symbol sequence and the previous segment text. During training, this context is the ground-truth previous segment \(R_{j-1}\); during inference, it is the model prediction \(\hat{R}_{j-1}\).

\(\mathrm{LLM}_A\) is prompted with the symbol sequence alone, while \(\mathrm{LLM}_B\) is prompted with both the previous segment text and the current symbol sequence.

Both models are trained with teacher-forced cross-entropy loss. Although symbolization is many-to-one, reconstruction remains effective because the model exploits sequential structure and contextual conditioning to resolve locally ambiguous symbols~\cite{loshchilov2017decoupled}. Additional training details are provided in Appendix~\ref{app:ml_details}.

\section{Evaluation}\label{sec:evaluation}
In this section, we outline the evaluation setup. Then we will explore the fidelity of the reconstruction attack along three primary dimensions: generality, robustness, and cost-effectiveness.

\subsection{Experiment Setup}\label{sec:setup}
\vspace{.3em}\noindent\textbf{Environment Hardware.}
To evaluate generality, we conduct experiments on two systems: a laptop and a desktop, both running Ubuntu 22.04 LTS. The laptop uses an Intel\textregistered{} Core i7-8650U CPU @ 1.90\,GHz with a 32\,KB L1 cache (8-way), 256\,KB L2 cache (4-way), and 8\,MB L3 cache (16-way). The desktop uses an 11th Gen Intel\textregistered{} Core\texttrademark{} i7-11700K CPU @ 3.60\,GHz with a 48\,KB L1 data cache (12-way), 32\,KB L1 instruction cache (8-way), 512\,KB L2 cache (8-way), and 16\,MB L3 cache (16-way), and is additionally equipped with NVIDIA RTX 3090 and RTX 3080 GPUs (CUDA 12.2).

\vspace{.3em}\noindent\textbf{Datasets \& Training.}
We use three publicly available LLM-generated datasets: Code-Alpaca~\cite{codealpaca}, UltraChat~\cite{ding2023ultrachat}, and ChatDoctor~\cite{li2023chatdoctor}, covering programming, medical queries, financial advice, and general knowledge. For multi-turn datasets, we retain only the first LLM response. Each dataset is randomly split into training, validation, and test sets using an 80\%/10\%/10\% ratio, with standard deduplication~\cite{lee2021deduplicating, brown2020language, chowdhery2023palm}. Further preprocessing details are provided in Appendix~\ref{sec:dataclean}.

Our evaluation additionally uses cache traces collected with the Prime+Probe procedure described in §\ref{sec:primeprobe_impl}. For each token, we collect 50 traces from repeated executions of the decode routine, split into 30 profiling traces for training and 20 held-out traces for evaluation. Thus, reconstruction is evaluated only on traces unseen during profiling.

For reconstruction, we fine-tune Flan-T5-XL for 5 epochs to obtain \( \mathrm{LLM}_A \) and \( \mathrm{LLM}_B \) across multiple settings. Training is performed on an NVIDIA RTX 6000 Pro GPU and takes up to 6 hours for \( \mathrm{LLM}_A \) and 4 days for \( \mathrm{LLM}_B \).

\vspace{.3em}\noindent\textbf{Metrics.}
We evaluate natural-language and code outputs separately, as reconstruction quality is captured differently for each. For UltraChat and ChatDoctor, we report normalized character-level Edit Distance (ED), where lower is better, ROUGE-1 F1~\cite{lin2004rouge} for unigram overlap, and cosine similarity $\phi$ between Sentence-Transformers \texttt{all-MiniLM-L6-v2} embeddings~\cite{reimers2019sentence}, where $\phi \in [-1,1]$ and higher values indicate stronger semantic similarity. We also use GPT-4.1-mini as an LLM-based judge and report Attack Success Rate (ASR) as the fraction of samples judged semantically equivalent, following prior work on LLM-based evaluation~\cite{gu2024llmjudge, zheng2023judging}.

For Code-Alpaca, lexical and embedding-based metrics do not reliably capture functional correctness. We therefore use the same judge to assess whether reconstructed and reference snippets implement the same functionality, allowing benign differences such as variable renaming, formatting, or alternative correct implementations~\cite{tong2024codejudge, crupi2025effectiveness}. Judge prompts are provided in Appendix Figure~\ref{fig:judge_prompts}.

\vspace{.3em}\noindent\textbf{Experiments.}
We evaluate the full reconstruction pipeline across multiple configurations, varying the target model between Phi-3-mini \cite{abdin2024phi}, which uses a 32k-token vocabulary, and Llama-3 \cite{grattafiori2024llama}, which uses a 128k-token vocabulary. We further vary the tokenization framework (\texttt{Llama.cpp} and HuggingFace \texttt{Transformers}), hardware platform (laptop and desktop), and dataset (UltraChat, Code-Alpaca, and ChatDoctor). For each configuration, each of the reconstruction models is trained on a disjoint training split and evaluated on unseen traces and responses, reporting results for both first-segment and full-response reconstruction.

\subsection{Attack Generality \label{sec:general_results}}

\begin{table*}[t]
\centering
\caption{Reconstruction performance for textual datasets, frameworks, hardware platforms, and target models. All metrics are reported as mean $\pm$ standard deviation. Higher is better for all metrics except ED ($\downarrow$).}
\label{tab:main_results}
\setlength{\tabcolsep}{3.2pt}
\renewcommand{\arraystretch}{1.08}
\scriptsize
\resizebox{\textwidth}{!}{%
\renewcommand{\arraystretch}{0.8}
\begin{tabular}{c c c c |c| ccc |c| ccc}
\toprule
\multirow{2}{*}{\textbf{Dataset}} &
\multirow{2}{*}{\textbf{Framework}} &
\multirow{2}{*}{\textbf{Hardware}} &
\multirow{2}{*}{\textbf{Target}} &
\multicolumn{4}{c|}{\textbf{First Segments}} &
\multicolumn{4}{c}{\textbf{Full Paragraphs}} \\
\cmidrule(lr){5-8} \cmidrule(lr){9-12}
& & & & \textbf{ASR} & \textbf{$\Phi$} & \textbf{R1-F1} & \textbf{ED} $\downarrow$
         & \textbf{ASR} & \textbf{$\Phi$} & \textbf{R1-F1} & \textbf{ED} $\downarrow$ \\
\midrule

\multirow{8}{*}{\rotatebox[origin=c]{90}{\textbf{UltraChat}}}
& \multirow{4}{*}{Llama.cpp}
  & \multirow{2}{*}{Laptop}
    & Phi-3-mini   & \textbf{57.51\%} & 69.24\% $\pm$ 31.59\% & 77.25\% $\pm$ 21.80\% & 77.97\% $\pm$ 20.63\%
            & \textbf{85.15\%} & 86.68\% $\pm$ 14.19\% & 82.70\% $\pm$ 9.69\% & 82.96\% $\pm$ 9.68\% \\
& & & Llama-3 & \textbf{39.71\%} & 54.53\% $\pm$ 33.54\% & 63.99\% $\pm$ 23.96\% & 65.72\% $\pm$ 22.18\%
            & \textbf{72.03\%} & 81.22\% $\pm$ 15.67\% & 72.56\% $\pm$ 10.28\% & 72.08\% $\pm$ 10.37\% \\
\cmidrule(lr){3-12}
& & \multirow{2}{*}{Desktop}
    & Phi-3-mini   & \textbf{61.55\%} & 71.62\% $\pm$ 30.98\% & 78.72\% $\pm$ 21.51\% & 79.35\% $\pm$ 20.32\%
            & \textbf{87.42\%} & 87.52\% $\pm$ 13.80\% & 83.02\% $\pm$ 9.54\% & 83.31\% $\pm$ 9.56\% \\
& & & Llama-3 & \textbf{44.52\%} & 59.41\% $\pm$ 33.28\% & 69.32\% $\pm$ 23.60\% & 70.18\% $\pm$ 22.10\%
            & \textbf{74.78\%} & 83.19\% $\pm$ 15.12\% & 76.52\% $\pm$ 10.21\% & 76.09\% $\pm$ 10.40\% \\
\cmidrule(lr){2-12}
& \multirow{4}{*}{HuggingFace}
  & \multirow{2}{*}{Laptop}
    & Phi-3-mini   & \textbf{58.77\%} & 69.43\% $\pm$ 31.18\% & 75.65\% $\pm$ 22.18\% & 77.05\% $\pm$ 20.66\%
            & \textbf{86.11\%} & 86.38\% $\pm$ 14.04\% & 80.23\% $\pm$ 9.69\% & 80.89\% $\pm$ 9.59\% \\
& & & Llama-3 & \textbf{25.50\%} & 40.08\% $\pm$ 33.26\% & 51.18\% $\pm$ 25.92\% & 54.21\% $\pm$ 22.90\%
            & \textbf{56.68\%} & 76.51\% $\pm$ 16.99\% & 64.78\% $\pm$ 11.48\% & 62.89\% $\pm$ 11.60\% \\
\cmidrule(lr){3-12}
& & \multirow{2}{*}{Desktop}
    & Phi-3-mini   & \textbf{55.22\%} & 66.58\% $\pm$ 32.70\% & 74.21\% $\pm$ 23.59\% & 75.40\% $\pm$ 21.86\%
            & \textbf{84.04\%} & 85.81\% $\pm$ 14.51\% & 80.82\% $\pm$ 10.17\% & 81.11\% $\pm$ 10.14\% \\
& & & Llama-3 & \textbf{43.02\%} & 57.70\% $\pm$ 33.45\% & 67.67\% $\pm$ 23.76\% & 68.87\% $\pm$ 22.23\%
            & \textbf{73.26\%} & 82.55\% $\pm$ 15.22\% & 75.17\% $\pm$ 10.18\% & 74.84\% $\pm$ 10.38\% \\

\midrule

\multirow{8}{*}{\rotatebox[origin=c]{90}{\textbf{ChatDoctor}}}
& \multirow{4}{*}{Llama.cpp}
  & \multirow{2}{*}{Laptop}
    & Phi-3-mini   & \textbf{84.67\%} & 84.38\% $\pm$ 23.27\% & 85.17\% $\pm$ 19.52\% & 86.74\% $\pm$ 17.67\%
            & \textbf{90.08\%} & 75.25\% $\pm$ 15.17\% & 63.89\% $\pm$ 12.62\% & 66.28\% $\pm$ 12.68\% \\
& & & Llama-3 & \textbf{48.67\%} & 56.40\% $\pm$ 27.75\% & 62.29\% $\pm$ 22.67\% & 65.56\% $\pm$ 20.50\%
            & \textbf{78.60\%} & 67.97\% $\pm$ 17.40\% & 66.37\% $\pm$ 14.09\% & 67.05\% $\pm$ 14.16\% \\
\cmidrule(lr){3-12}
& & \multirow{2}{*}{Desktop}
    & Phi-3-mini   & \textbf{63.89\%} & 64.00\% $\pm$ 29.22\% & 62.02\% $\pm$ 25.86\% & 62.71\% $\pm$ 22.43\%
            & \textbf{91.27\%} & 74.24\% $\pm$ 17.35\% & 70.55\% $\pm$ 14.22\% & 71.01\% $\pm$ 13.69\% \\
& & & Llama-3 & \textbf{51.58\%} & 59.14\% $\pm$ 28.41\% & 65.53\% $\pm$ 23.52\% & 68.27\% $\pm$ 21.19\%
            & \textbf{81.24\%} & 69.74\% $\pm$ 17.76\% & 69.08\% $\pm$ 14.62\% & 69.60\% $\pm$ 14.57\% \\
\cmidrule(lr){2-12}
& \multirow{4}{*}{HuggingFace}
  & \multirow{2}{*}{Laptop}
    & Phi-3-mini   & \textbf{66.98\%} & 67.62\% $\pm$ 26.56\% & 69.12\% $\pm$ 22.10\% & 72.44\% $\pm$ 19.73\%
            & \textbf{92.85\%} & 75.97\% $\pm$ 16.54\% & 73.76\% $\pm$ 13.10\% & 75.56\% $\pm$ 12.65\% \\
& & & Llama-3 & \textbf{34.23\%} & 47.50\% $\pm$ 27.93\% & 52.21\% $\pm$ 25.55\% & 57.06\% $\pm$ 22.33\%
            & \textbf{59.76\%} & 61.02\% $\pm$ 17.72\% & 56.10\% $\pm$ 16.06\% & 56.30\% $\pm$ 16.20\% \\
\cmidrule(lr){3-12}
& & \multirow{2}{*}{Desktop}
    & Phi-3-mini   & \textbf{67.26\%} & 68.80\% $\pm$ 26.76\% & 72.26\% $\pm$ 22.14\% & 74.70\% $\pm$ 19.80\%
            & \textbf{92.80\%} & 76.62\% $\pm$ 16.67\% & 75.40\% $\pm$ 13.60\% & 76.56\% $\pm$ 13.20\% \\
& & & Llama-3 & \textbf{51.85\%} & 59.59\% $\pm$ 27.97\% & 65.95\% $\pm$ 22.38\% & 68.38\% $\pm$ 20.49\%
            & \textbf{80.61\%} & 69.73\% $\pm$ 17.61\% & 69.11\% $\pm$ 14.26\% & 69.72\% $\pm$ 14.30\% \\

\bottomrule
\end{tabular}%
}
\end{table*}
\begin{table}[]
\centering
\caption{Code-Alpaca reconstruction performance (ASR) for first-segment and full-output reconstruction.}
\label{tab:code_results}
\setlength{\tabcolsep}{3pt}
\renewcommand{\arraystretch}{0.8}
\footnotesize
\begin{tabular}{l l l c c}
\toprule
\textbf{Framework} & \textbf{Hardware} & \textbf{Target} & \textbf{First} & \textbf{Full} \\
\midrule
\multirow{4}{*}{Llama.cpp}
& \multirow{2}{*}{Laptop}
& Phi-3-mini   & 90.17\% & 95.87\% \\
& & Llama-3 & 50.13\% & 49.43\% \\
\cmidrule(lr){2-5}
& \multirow{2}{*}{Desktop}
& Phi-3-mini   & 87.65\% & 90.56\% \\
& & Llama-3 & 81.55\% & 86.86\% \\
\midrule
\multirow{4}{*}{HuggingFace}
& \multirow{2}{*}{Laptop}
& Phi-3-mini   & 85.19\% & 88.44\% \\
& & Llama-3 & 64.69\% & 71.35\% \\
\cmidrule(lr){2-5}
& \multirow{2}{*}{Desktop}
& Phi-3-mini   & 84.92\% & 89.71\% \\
& & Llama-3 & 79.19\% & 84.72\% \\
\bottomrule
\end{tabular}
\end{table}

In this section, we evaluate the attack's performance in reconstructing both the initial segment and full paragraphs across all configurations. We distinguish between first-segment and full-response reconstruction, as the initial portion often reveals the core confidential topic.

\begin{figure}[t]
\begin{tcolorbox}[title = Attack Examples - Ultrachat,width=\columnwidth]
\setstretch{0.9}

\underline{Judge: Success \hspace{2.2em} $\phi: 1.00$ \hspace{2.2em} R1: $1.00$ \hspace{2.2em} ED: $0.00$}
\vspace{.4em}\\
\small
\textbf{Ref:} Yes, incorporating daily exercise into your routine can improve your mental health in several ways. Exercise can help reduce symptoms of depression and anxiety
\vspace{.15em}

\textbf{Pred:} Yes, incorporating daily exercise into your routine can improve your mental health in several ways. Exercise can help reduce symptoms of depression and anxiety
\normalsize

\vspace{.8em}
\underline{Judge: Success \hspace{1.9em} $\phi: 0.835$ \hspace{1.9em} R1: $0.70$ \hspace{1.9em} ED: $0.296$}
\vspace{.4em}\\
\small
\textbf{Ref:} There are several types of therapy that are effective for treating anxiety and depression: 1. \textcolor{black}{\textbf{Cognitive Behavioral Therapy (CB}} 
\vspace{.15em}

\textbf{Pred:} There are several types of therapy that are effective for treating anxiety and depression \textcolor{BrickRed}{\textbf{disorders (ADHD) by combining counseling and therapy}}
\normalsize

\vspace{.8em}
\underline{Judge: Success \hspace{1.9em} $\phi: 0.699$ \hspace{1.9em} R1: $0.837$ \hspace{1.9em} ED: $0.153$}
\vspace{.4em}\\
\small
\textbf{Ref:} \textcolor{black}{\textbf{Both cardio and strength training}} can be effective in \textcolor{black}{\textbf{weight loss strategies}}. Cardio exercises such as running, cycling, and swimming can help
\vspace{.15em}

\textbf{Pred:} \textcolor{BrickRed}{\textbf{Cross-training and strength training}} can be effective in \textcolor{BrickRed}{\textbf{weight loss surgery}}. Cardio exercises such as running, cycling, and swimming often help
\normalsize

\vspace{.8em}
\underline{Judge: Success \hspace{1.9em} $\phi: 0.451$ \hspace{1.9em} R1: $0.625$ \hspace{1.9em} ED: $0.336$}
\vspace{.4em}\\
\small
\textbf{Ref:} \textcolor{black}{\textbf{Fibromyalgia and Chronic Fatigue Syndrome (CFS)}} share some common symptoms, such as fatigue, sleep problems,
\vspace{.15em}

\textbf{Pred:} \textcolor{BrickRed}{\textbf{Cyanobacteria and fungi}} do not always and everywhere share some common symptoms, such as fatigue, sleep problems,
\normalsize

\vspace{.8em}
\underline{Judge: Failure \hspace{1.9em} $\phi: 0.532$ \hspace{1.9em} R1: $0.652$ \hspace{1.9em} ED: $0.287$}
\vspace{.4em}\\
\small
\textbf{Ref:} Studies have shown that \textcolor{black}{\textbf{drinking beer in moderation}} may have some cardiovascular health benefits, particularly for those at risk of heart disease.
\vspace{.15em}

\textbf{Pred:} Studies have shown that \textcolor{BrickRed}{\textbf{athletes eating disorders often}} may have some cardiovascular health benefits, particularly at the intermediate levels of heart disease.
\normalsize



\end{tcolorbox}
\vspace{-1em}
\caption{Examples of successful and unsuccessful attacks on $R_0$ in UltraChat.}
\label{fig:attack_examples}
\end{figure}

The full results are presented in Table~\ref{tab:main_results} for textual datasets and in Table~\ref{tab:code_results} for code datasets, with real reconstruction examples in Figure \ref{fig:attack_examples}. Across all settings, reconstruction achieves high semantic fidelity despite ambiguous token-level signals. For full paragraphs, UltraChat reaches ASR of $\sim$56--87\% with semantic similarity $\phi$ of $\sim$76--87\%, while ChatDoctor achieves substantially higher ASR of $\sim$78--93\% with $\phi$ of $\sim$61--76\%. Despite weak per-token signals, reconstructions often preserve subtopics and local structure even when individual words are incorrect.

This semantic robustness stems from sequence-level modeling: although individual symbols are noisy and ambiguous, the model leverages context across the sequence to resolve uncertainty, enabling accurate semantic reconstruction even when lexical similarity degrades.

To further characterize reconstruction fidelity beyond broad semantic recovery, we examine the fraction of outputs achieving very high semantic similarity ($\phi \geq 0.9$). For first-segment reconstruction, this rate reaches up to $62.92\%$ on ChatDoctor and $44.62\%$ on UltraChat. Even for full responses, up to $58.48\%$ of UltraChat reconstructions achieve $\phi \geq 0.9$. These results show that the attack can recover outputs with high fidelity, rather than merely identifying their broad topic. Appendix~\ref{appendix:high-fidelity} provides the full breakdown.

\begin{tcolorbox}[colback=observationgreen,colframe=green!80!black,fonttitle=\bfseries,boxsep=1mm, bottom=0mm, left=1mm, right=1mm, top=0mm]
\textbf{Insight.} Semantic recovery makes the attack effective even without exact token inference, while the high-fidelity results show that the leakage can extend well beyond topic-level information to closely reconstructed outputs.
\end{tcolorbox}

Performance is consistent across hardware platforms and inference frameworks, indicating that the attack does not depend on specific microarchitectural properties. Instead, it relies on stable decode-table access patterns within a given deployment, together with effective profiling to learn the corresponding mappings.

Performance differences are primarily driven by the target model and dataset. Llama consistently underperforms Phi, which we attribute to its larger vocabulary ($\times4$) and the resulting increase in cache-set collisions. Additionally, more structured datasets such as ChatDoctor and Code-Alpaca yield higher performance than more diverse datasets like UltraChat, suggesting that constrained token distributions reduce reconstruction ambiguity.

\subsection{Attack Robustness}
All experiments so far use real cache traces and therefore already include natural measurement noise. To further study robustness, we inject additional synthetic noise, with probability \(p\), into the symbolic token sequences produced by the clustering stage. Specifically, we introduce deletions, insertions, and substitutions in the clustering output. In our reconstruction pipeline, deletions correspond to false negatives in the auxiliary detection side channel, where a \texttt{decode} event is missed and no trace is collected. Insertions correspond to false positives, where a spurious cache hit is detected despite no true \texttt{decode} event. Substitutions represent clustering misclassifications, where a collected cache trace is assigned the wrong symbolic token. This form of synthetic corruption is commonly used in prior side-channel reconstruction work to evaluate robustness to measurement errors~\cite{gao2025iknowwhatyousaid, yuan2022automated, zhang2024r+}.

For each value of \(p\), we evaluate our base reconstruction model \(\mathrm{LLM}_A\), trained on standard UltraChat traces collected on the laptop using \texttt{Llama.cpp} with Phi-3 as the target model. The model is trained only on the original, uncorrupted traces. During evaluation, each token in the clustering output is independently corrupted with probability \(p\), with the corruption type chosen uniformly among deletion, insertion, and substitution. This allows us to systematically measure how reconstruction quality degrades as noise increases.

\begin{figure}[]
    \centering
    \includegraphics[width=\columnwidth]{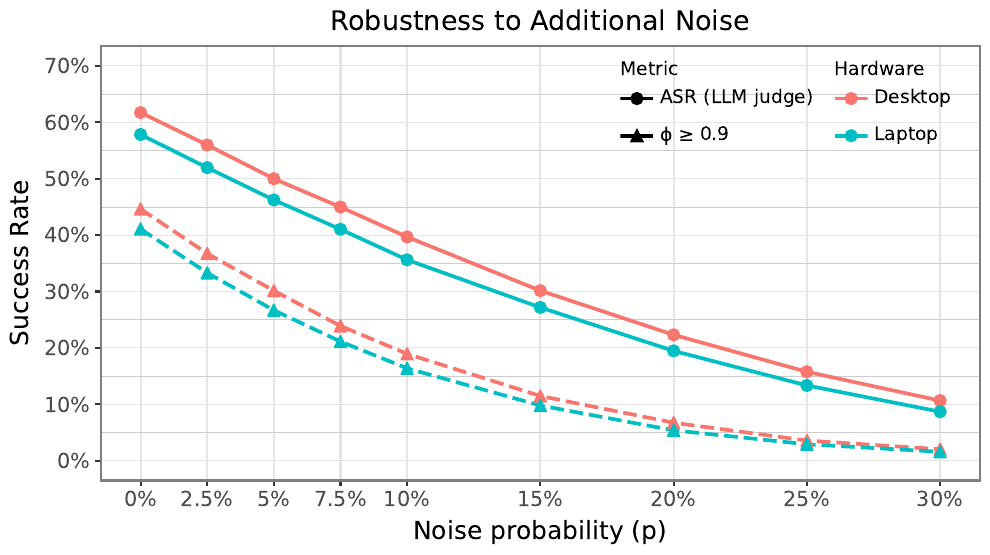}
    \caption{Robustness to additional noise injected into the reconstruction pipeline. We report, for each $p$, the base ASR and the rate of high-fidelity predictions with $\phi \geq 0.9$ for both hardware settings.}
    \label{fig:noise_robustness}
\end{figure}

As shown in Figure~\ref{fig:noise_robustness}, performance gradually degrades as \(p\) increases. Even at \(p = 0.15\), \(\mathrm{LLM}_A\) still attains an ASR of \(\sim30\%\), with roughly \(10\%\) of reconstructions achieving high semantic fidelity (\(\phi > 0.9\)). These results show that reconstruction remains robust under substantial injected noise.

\subsection{Query Performance Tradeoff \label{sec:query_tradeoff}}

The profiling phase is critical to the reconstruction attack, as it establishes the mapping between cache activity and output tokens. In practice, profiling presents a tradeoff between the number of queries and how distinctive they are. We therefore evaluate two approaches: \textit{targeted profiling}, which deliberately elicits selected tokens with few queries, and \textit{benign profiling}, which uses only ordinary conversational queries at the cost of a larger budget.

\begin{figure}[ht]
    \centering
    \includegraphics[width=0.8\columnwidth]{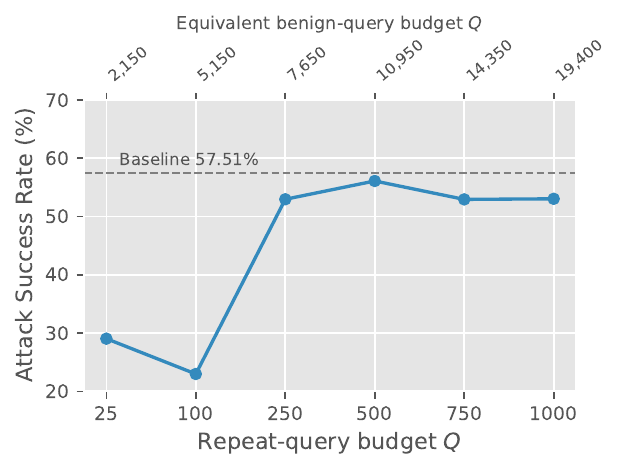}
    \caption{ASR as a function of targeted and equivalent benign profiling query budgets on Phi-3-mini.}
    \label{fig:asr_vs_queries}
\end{figure}

We evaluate this tradeoff using the laptop, UltraChat, \texttt{Llama.cpp}, and Phi-3 setting. In targeted profiling, the attacker issues queries of the form \textit{``repeat the following sequence 5 times: $\langle$sequence of 10 tokens$\rangle$''}, obtaining five traces per selected token. Tokens are chosen by frequency in the UltraChat training split. For each targeted-query budget (\(Q_t\)), we construct an equivalent profile using ordinary UltraChat-style queries until they provide the same profiled token set and trace coverage. Figure~\ref{fig:asr_vs_queries} reports the resulting ASR, with the corresponding benign-query budget (\(Q_b\)) shown on the upper axis.

Targeted profiling is highly query-efficient: with only \(Q_t=250\) queries, the attack exceeds 55\% ASR and completes within roughly 6 minutes of standard inference on the tested hardware. Distinctive repetition queries, however, are not required. An equivalent profile can be obtained with approximately \(Q_b=7{,}650\) ordinary conversational queries, requiring roughly 3 hours and achieving about 53\% ASR. With \(Q_b=10{,}950\) benign queries, ASR reaches approximately 56\%, close to the unlimited-profile baseline of 57.51\%. Further increases provide little additional benefit, with performance remaining around 53--56\%.

These results show that profiling can be adapted to the attacker's constraints: targeted queries enable rapid profiling, while ordinary conversational queries retain near-baseline reconstruction performance with a larger budget.

\subsection{Hardware Evaluation} \label{sec:hardware_evaluation}

To evaluate the attack on recent hardware, we repeat the first-segment evaluation on a 13th Gen Intel Core i9-13950HX using Llama.cpp with Phi-3-mini and Llama-3 on UltraChat. Modern processors employ hardware prefetchers that can introduce additional cache accesses, interfering with cache-based side-channel measurements~\cite{van2018foreshadow,wang2019papp}. As shown in Table~\ref{tab:modern_hardware_evaluation}, performance degrades relative to our earlier results (Table~\ref{tab:main_results}), but the attack remains effective, demonstrating that the leakage persists on modern Intel hardware despite these effects.

\begin{table}[ht]
\centering
\caption{Reconstruction performance (first-segments) on a 13th Gen Intel CPU.}
\small
\renewcommand{\arraystretch}{0.85}
\label{tab:modern_hardware_evaluation}
\begin{tabular}{lcccc}
\toprule
\textbf{Target} & \textbf{ASR} & $\boldsymbol{\phi > 0.5}$ & \textbf{R1-F1} & \textbf{ED} $\downarrow$ \\ 
\midrule
Phi-3-mini  & 38.40\%  & 52.12\% & 56.97\% & 61.07\% \\
Llama-3     & 31.97\%  & 46.86\% & 53.87\% & 57.97\% \\
\bottomrule
\end{tabular}
\end{table}

\subsection{End-to-End Attack on OpenClaw}
So far, our evaluation relied on dataset-aligned traces, where cache measurements were collected and matched to pre-existing text corpora. While this isolates the reconstruction pipeline, it does not capture a fully realistic deployment. We now evaluate the attack end-to-end on a live system by performing profiling, training, and reconstruction directly on an OpenClaw agent \cite{openclaw} running a local Phi-3-mini model \cite{abdin2024phi}, without any synthetic alignment. OpenClaw is a general-purpose open-source LLM agent framework for local deployment, enabling persistent assistants that manage personal workflows and interact with local resources, and is among the most popular such systems by GitHub stars (Figure~\ref{fig:vuln_mapping}).

We use a subset of UltraChat prompts \((n = 25,000)\) as inputs to the agent, issuing them and collecting traces during real execution. We evaluate reconstruction on the first response segments only, as these are sufficient to recover the main topic of the output, as discussed in Section~\ref{sec:general_results}. This reflects a practical deployment in which an LLM-based agent processes user queries over time.

We report reconstruction performance using ASR and mean semantic similarity $\phi$. Since OpenClaw relies on \texttt{Llama.cpp} for local inference (Figure~\ref{fig:vuln_mapping}), we compare against the corresponding controlled setting (Llama.cpp, Phi, Desktop, UltraChat), which achieves 61.55\% ASR and 71.62\% $\phi$ ($\pm$ 30.98\%). On OpenClaw, the attack achieves 30.12\% ASR with a mean $\phi$ of \(42.23\% \pm 20.28\%\). Performance degrades due to additional end-to-end noise, including timing variability and imperfect alignment. Nevertheless, the attack remains effective, showing that reconstruction transfers to a real-world deployment.

\section{Mitigation Strategies}\label{sec:mitigations}  

In this section, we discuss potential mitigations and their practical limitations in the context of our threat model.

\vspace{.3em}\noindent\textbf{Decode Table Randomization.}
As described in Section~\ref{sec:threat_model}, our attack relies on a stable mapping between token identifiers and their memory footprint during the lifetime of the victim process. A natural mitigation is therefore to randomize the in-memory layout of the decode table during execution, for example, by shuffling token placements or permuting internal indexing structures. This changes the mapping between token IDs and cache sets, invalidating previously collected traces and disrupting the learned leakage model. However, an active adversary may re-profile the victim after each re-randomization event. Thus, randomization primarily increases the attack cost rather than eliminating the underlying leakage channel.

\vspace{.3em}\noindent\textbf{Process Lifetime Reduction.}
Our reconstruction model is process-specific and does not transfer across independent instances due to ASLR, heap allocation differences, and randomized data structure layouts. Periodic restarts or short-lived workers can therefore limit the time available for profiling. However, frequent model initialization increases latency and reduces responsiveness. Moreover, this mitigation does not eliminate the leakage, as an adversary can re-profile each new instance.

\vspace{.3em}\noindent\textbf{Microarchitectural and OS-level Defenses.}
Disabling SMT, such as Intel's Hyper-Threading~\cite{marr2002hyperthreading}, prevents an attacker from sharing a physical core with the victim and therefore blocks L1 Prime+Probe~\cite{tromer2010efficient}. However, disabling SMT incurs a significant performance cost (Section~\ref{sec:threat_model}), making it unattractive for resource-constrained local LLM deployments.

Cache partitioning, such as page coloring~\cite{ye2014coloris}, and hardware mechanisms that randomize cache indexing or eviction~\cite{qureshi2018ceaser} can also reduce the available signal, although profiling may partially adapt. These defenses are generally not enabled on consumer systems due to limited support and performance or deployment overhead~\cite{mushtaq2020winter}, making them more suitable for multi-tenant or high-assurance environments.

\section{Related Works}\label{sec:related_works}  

\vspace{.3em}\noindent\textbf{Privacy Attacks on Language Models.} LLMs expose privacy risks at the application and network levels. Deployment optimizations such as KV-cache reuse across tokens or sessions have been exploited to infer user prompts and model outputs through fine-grained decoding timing~\cite{song2024early} or probing shared KV-cache entries in multi-tenant settings~\cite{wu2025know}. At the network level, observing the length and structure of encrypted token streams can leak model outputs~\cite{weiss2024your}. Other works show that adversarial queries can elicit hidden system prompts or reconstruct prior conversations, revealing sensitive context through repeated interaction~\cite{hui2024pleak,yang2024prsa,chu2024reconstruct,debenedetti2024privacy}. Together, these studies highlight non-hardware side channels that persist even under strict access isolation.

\vspace{.3em}\noindent\textbf{Hardware Side-Channel Attacks on Deep Learning.} Hardware side channels have been used to infer neural-network inputs and outputs through power analysis, electromagnetic emanations, timing variations, or memory bus contention~\cite{liu2023side, shukla2023whispering}. Other studies extract model architectures and weights, including layer parameters, connectivity, and numerical values~\cite{gao2024deeptheft, batina2019csi, horvath2024sok}. These attacks exploit structured, data-dependent behavior in DNN execution.

Cache-based attacks pursue similar goals. Prior work shows that cache activity can leak sensitive user inputs~\cite{yuan2022automated, zhang2024r+, yuan2021private} or reconstruct model architectures and weights across deployment settings \cite{liu2024deepcache, yan2020cache, wang2022stealthy}. Autoregressive LLMs, however, exhibit fundamentally different execution characteristics: they generate outputs token-by-token, rely on large vocabulary lookups, and induce input-dependent, non-repetitive memory access patterns. These properties make prior techniques, which assume static or highly regular execution, difficult to apply directly.

\vspace{.3em}\noindent\textbf{Hardware Side-Channel Attacks on LLMs.}

Recent work has demonstrated hardware side-channel leakage from local LLM inference, but existing attacks rely on leakage sources that are present only under particular model or deployment conditions. Table~\ref{tab:assumptions_comparison} summarizes these requirements and contrasts them with ours.

One line of work exploits shared model data, particularly token embeddings. Adiletta and Sunar~\cite{adiletta2025spill} use Flush+Reload on embedding vectors exposed through CUDA unified CPU/GPU memory to leak tokens and partially recover API keys, requiring a non-default CUDA memory configuration. Gao et al.~\cite{gao2025iknowwhatyousaid} similarly monitor embedding-table accesses through shared model pages, requiring both the model data to be shared (e.g., through \texttt{mmap} or page deduplication) and the embedding lookup to execute on the CPU. When the model fits entirely in GPU memory, as is common in local inference, the embedding layer remains GPU-resident, and this leakage channel is absent. In contrast, our attack targets CPU-side detokenization after token generation, avoiding these model-data visibility and placement requirements altogether.

\begin{table}[t]
\centering
\caption{Comparison of deployment assumptions across local LLM side-channel attacks. "Special" denotes non-standard shared resources, while "standard" denotes commonly shared resources in default deployments.}
\label{tab:assumptions_comparison}
\footnotesize
\begin{tabular}{l|cc|cc}
\toprule
& \multicolumn{2}{c|}{\textbf{Model Assumptions}} 
& \multicolumn{2}{c}{\textbf{System Assumptions}} \\
\textbf{Attack} & \textbf{Weights} & \textbf{Arch} & \textbf{SMT} & \textbf{Shared Resources} \\ 
\midrule
IKWYS~\cite{gao2025iknowwhatyousaid} & CPU & None & None & Model data; special \\
Spill the Beans~\cite{adiletta2025spill} & CPU & None & None & Model data; special \\
MoEcho-GPU~\cite{ding2025moecho} & None & MoE & None & GPU counters; special \\
MoEcho-CPU~\cite{ding2025moecho} & None & MoE & Yes & Expert pages; special \\
\textbf{Ours} & \textbf{None} & \textbf{None} & \textbf{Yes} & \textbf{Code only; standard} \\
\bottomrule
\end{tabular}
\end{table}

Another line of work focuses on \textit{specific model architectures}. Ding et al.~\cite{ding2025moecho} show that MoE models leak expert activation patterns, enabling prompt inference and response reconstruction. Their attack variants additionally rely on specialized observability, such as GPU performance counters or shared expert data pages. More fundamentally, however, the leakage source is the MoE routing mechanism itself. As a result, the attack does not extend to standard dense models, such as common Llama-style deployments. In contrast, our attack targets detokenization, a model-agnostic step performed after each generated token.

This different leakage source also creates a substantially weaker observation channel. Prior embedding attacks observe shared data at cache-line granularity~\cite{adiletta2025spill, gao2025iknowwhatyousaid}, while MoE attacks exploit large, structured expert-execution footprints~\cite{ding2025moecho}. Network attacks similarly expose stable observables such as response size or latency~\cite{weiss2024your, wu2025know, song2024early}. A detokenization event, by contrast, is short-lived and touches few cache lines. Prime+Probe further collapses many token-dependent accesses onto the same cache sets. Individual traces therefore do not directly reveal token identities. We overcome this by mapping noisy traces to stable symbolic representations and using language context to resolve the resulting ambiguity.

Overall, prior attacks expose important leakage channels in local LLM inference, but their applicability is tied to either model-internal data exposure or architecture-specific execution behavior. To our knowledge, the decode table has not previously been examined as a leakage source. Our work shifts the attack surface to detokenization, a default component of the generation pipeline, enabling cache-based output reconstruction across a broader range of local LLM deployments than previously demonstrated.

\section{Conclusion}\label{sec:conclusion}

We presented a new cache-based attack that reconstructs the text generated by locally hosted LLMs by observing CPU cache activity during detokenization. By targeting the decode table, our attack avoids the deployment-specific assumptions required by prior work and instead exploits a leakage source that is present in default LLM inference pipelines. We showed how to combine Flush+Reload for decode detection, Prime+Probe for cache-trace capture, and a clustering-and-language-model pipeline for text reconstruction from noisy observations. Our findings identify detokenization as a practical and widespread leakage surface in local LLM serving and highlight the need for stronger isolation and defenses in privacy-sensitive deployments.

\section{Acknowledgments}
The authors would like to thank Lital Badash from Microsoft for her help in managing the project.

\bibliographystyle{ACM-Reference-Format}
\bibliography{references}

\appendix
\section{Open Science}
We use publicly available models, frameworks, and datasets throughout this work, including Phi-3-mini, Llama-3, \texttt{Llama.cpp}, HuggingFace \texttt{Tokenizers}, UltraChat, ChatDoctor, and Code-Alpaca.

We do not provide exploit artifacts or components that could directly enable misuse against real deployments. To support reproducibility, we present our code, reconstruction pipelines, and evaluation scripts at the following link: \url{https://github.com/royweiss1/Detokenization-Leaks}.

\section{Ethical Considerations}
All experiments are conducted on local machines using open-source models and controlled datasets. No real user data or human subjects are involved. The identified vulnerabilities have been responsibly disclosed to relevant parties, including \texttt{HuggingFace} and \texttt{Llama.cpp} maintainers.

\section{Data Cleaning \label{sec:dataclean}}
To reduce redundancy in the training data, we applied an 8-gram deduplication procedure to the user prompts, following the approach adopted in PaLM and GPT-3 training \cite{brown2020language, chowdhery2023palm}. A training sample is removed if its prompt shares its first 8 words with any previously seen prompt, or if more than 70\% of its 8-grams overlap with those of any previously seen prompt. This deduplication criterion has also been employed in prior work \cite{gao2025iknowwhatyousaid}. Additionally, we applied further cleaning to the UltraChat dataset to remove empty or generic responses (e.g., ``As an AI, I can't help with that''). See Table \ref{tab:dataset_cleaning} for dataset statistics before and after filtering.

\begin{table}[htbp]
\centering
\small
\begin{tabular}{@{}lrrc@{}}
\toprule
\textbf{Dataset} & \textbf{Before} & \textbf{After} & \textbf{Cleaned (\%)} \\ \midrule
Code-Alpaca   & 121,959  & 98,540   & 18.8\%  \\
ChatDoctor    & 112,165  & 96,084   & 14.3\%   \\
UltraChat     & 577,747  & 358,545  & 37.9\%  \\ \bottomrule
\end{tabular}%
\caption{Dataset sizes before and after cleaning, with percentage removed.}
\label{tab:dataset_cleaning}
\end{table}

\section{LLM Prompts \label{appendix:llm_train_prompt}}
Suppose 
\(R_0=\) ``\textit{Based on the medical information you provided, here is an explanation of the results.}'' and 
\(R_1=\) ``\textit{The report indicates that the tiredness you feel is due to stress and recommends follow-up testing.}'' 
The training inputs take the form:

\begin{tcolorbox}[title=Training Prompts, width=\columnwidth]
\small

\textbf{\(\mathrm{LLM}_A\) (first segment)}

Translate the symbolic tokens to English.

\textbf{Symbolic Tokens:}

$\langle \mathrm{SYM}_{12} \rangle\ 
\langle \mathrm{SYM}_{7} \rangle\ 
\langle \mathrm{SYM}_{31} \rangle\ 
\langle \mathrm{SYM}_{4} \rangle\ 
\langle \mathrm{SYM}_{18} \rangle\ 
\langle \mathrm{SYM}_{12} \rangle\ 
\langle \mathrm{SYM}_{3} \rangle\ \dots$

\vspace{0.6em}

\textbf{\(\mathrm{LLM}_B\) (middle segment)}

Translate the symbolic tokens to English, given the context.

\textbf{Context (reconstructed):}

\textit{Based on the medical \textcolor{BrickRed}{\textbf{report}} you \textcolor{BrickRed}{\textbf{gave}}, here is an explanation of the \textcolor{BrickRed}{\textbf{diagnosis}}.}

\textbf{Symbolic Tokens:}

$\langle \mathrm{SYM}_{5} \rangle\ 
\langle \mathrm{SYM}_{22} \rangle\ 
\langle \mathrm{SYM}_{9} \rangle\ 
\langle \mathrm{SYM}_{41} \rangle\ 
\langle \mathrm{SYM}_{5} \rangle\ 
\langle \mathrm{SYM}_{14} \rangle\ 
\langle \mathrm{SYM}_{27} \rangle\ \dots$

\end{tcolorbox}

\begin{figure}[htbp]
\centering
\small
\begin{tcolorbox}[title=Reconstruction Judge Prompt]
\textbf{System Prompt (UltraChat): } "You are a privacy-leakage evaluator for reconstructed responses from side-channel observations; judge whether the prediction reveals meaningful and accurate information about the reference rather than quality, and answer Yes if it captures the same topic, intent, structure, style, or key points even if partial or noisy, otherwise No; output only Yes or No." \\

\textbf{System Prompt (ChatDoctor): } "You are a privacy-leakage evaluator for reconstructed medical responses from side-channel observations; judge whether the prediction reveals meaningful and accurate clinical information about the reference rather than quality or safety, and answer Yes if it reflects the same specific symptoms, conditions, diagnoses, treatments, medications, reasoning, or structured advice even if partial or noisy, otherwise No; output only Yes or No" \\

\textbf{System Prompt (Code-Alpaca): } "You are a privacy-leakage evaluator for reconstructed code from side-channel observations; judge whether it reveals meaningful and accurate information about the original code rather than correctness, and answer Yes if it reflects the same general topic, structure, frameworks, APIs, algorithms, or templates even if partial or noisy, otherwise answer No; output only Yes or No." \\

\textbf{User Prompt: } \\
Reference: <Original> \\
Prediction: <Reconstructed> \\
\end{tcolorbox}
\caption{LLM judge prompts for reconstruction evaluation.}
\label{fig:judge_prompts}
\end{figure}

\section{Additional Modeling Details}
\label{app:ml_details}

\noindent\textbf{Trace preprocessing and representation.}
Each trace is represented as a 64-dimensional latency vector. We subtract the per-trace mean and apply feature-wise standardization over the profiling set. During training, batches are constructed by sampling token IDs and drawing multiple traces per token, producing multiple positives per token alongside cross-token negatives for contrastive learning.

\noindent\textbf{Clustering and symbolization.}
Following encoding, we compute a single centroid per token by averaging the embeddings of its profiling traces. We then cluster these centroids using K-means with cosine similarity. At inference, each trace is encoded and assigned to its nearest cluster center to obtain a symbolic token. This two-stage approach—encoding followed by K-means—consistently outperforms applying K-means directly to the raw traces, with particularly pronounced gains on lower-performing tokens (2$\times$ improvement on the worst-performing 10\% of tokens). We attribute this to the encoder's ability to suppress noise and amplify token-discriminative structure in the latent space.

\noindent\textbf{Clustering quality.}
We evaluate clustering quality on held-out traces (20 test traces and 30 training traces) by measuring, for each token, the fraction of its traces assigned to the same cluster as its centroid, then averaging across tokens. This yields a mean clustering accuracy of approximately 75\% across all trace datasets.

\noindent\textbf{Reconstruction and decoding.}
Training examples for \(\mathrm{LLM}_B\) pair the preceding segment text as context with the current symbol sequence and target text. Unless stated otherwise, decoding is performed with beam search (4 beams, no sampling, maximum generation length of 38 tokens).

\section{K Ablation Study}\label{app:k-ablation}
\begin{figure}[th]
    \centering
    \includegraphics[width=\linewidth]{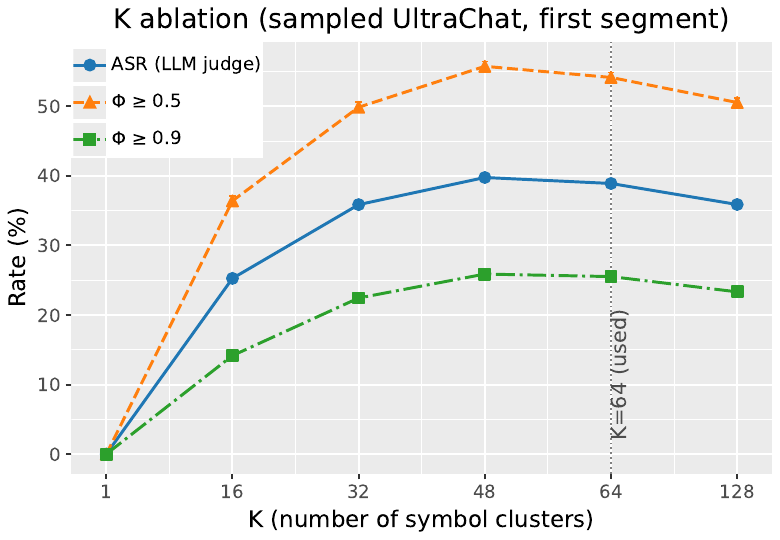}
    \caption{Reconstruction fidelity across different numbers of symbol clusters $K$ on UltraChat first-segment reconstruction.}
    \label{fig:k-ablation}
\end{figure}

We sweep $K \in \{1, 16, 32, 48, 64, 128\}$ on UltraChat (\texttt{Llama.cpp}, Desktop, Phi-3-mini), training a separate pipeline for each $K$ on $100{,}000$ training samples and evaluating on $5{,}000$ held-out samples. As shown in Figure~\ref{fig:k-ablation}, $K=1$ collapses to near-zero ASR, confirming the model cannot reconstruct text from an uninformative symbol alphabet. ASR rises sharply and peaks around $K=48$--$64$, then declines as larger $K$ fragments the per-symbol training signal. $K=48$ and $K=64$ are within noise of each other; we use $K=64$.

\section{High-Fidelity Reconstructions} \label{appendix:high-fidelity}

While the main evaluation characterizes reconstruction quality across the full test set, we additionally examine the upper end of the fidelity distribution. Table~\ref{tab:high-fidelity-breakdown} reports the fraction of reconstructions with high semantic similarity ($\phi\geq 0.9$), high lexical overlap (R1 $\geq 0.9$), and low normalized edit distance (ED $\leq 0.1$), together with their exact-match cases.

High-fidelity recovery is most pronounced for the first segments. For ChatDoctor, Phi-3-mini with Llama.cpp on the laptop reaches $60.43\%$ at ED $\leq 0.1$, including $36.06\%$ exact matches. For UltraChat, Phi-3-mini with Llama.cpp on the desktop reaches $40.43\%$ and $18.36\%$, respectively. High-fidelity recovery also occurs for complete responses; for example, the same UltraChat desktop configuration achieves $24.83\%$ at ED $\leq 0.1$. These results complement the aggregate semantic metrics by quantifying cases in which the reconstructed output closely matches the original text.

\begin{table*}[t]
\centering
\caption{Reconstruction accuracy across hardware, inference framework, target model, and segmentation configurations. Results are reported for high-fidelity reconstruction thresholds and exact matches under $\phi$, ROUGE-1 (R1), and edit-distance (ED) metrics.}
\label{tab:high-fidelity-breakdown}
\setlength{\tabcolsep}{3.2pt}
\renewcommand{\arraystretch}{1.08}
\scriptsize
\resizebox{\textwidth}{!}{%
\renewcommand{\arraystretch}{0.8}
\begin{tabular}{c c c c |cccccc|cccccc}
\toprule
\multirow{2}{*}{\textbf{Dataset}} &
\multirow{2}{*}{\textbf{Framework}} &
\multirow{2}{*}{\textbf{Hardware}} &
\multirow{2}{*}{\textbf{Target}} &
\multicolumn{6}{c|}{\textbf{First Segments}} &
\multicolumn{6}{c}{\textbf{Full Paragraphs}} \\
\cmidrule(lr){5-10} \cmidrule(lr){11-16}
& & & & \boldmath{$\phi{\geq}0.9$} & \boldmath{$\phi{=}1$} & \textbf{R1$\geq$0.9} & \textbf{R1$=$1} & \textbf{ED$\leq$0.1} & \textbf{ED$=$0} & \boldmath{$\phi{\geq}0.9$} & \boldmath{$\phi{=}1$} & \textbf{R1$\geq$0.9} & \textbf{R1$=$1} & \textbf{ED$\leq$0.1} & \textbf{ED$=$0} \\
\midrule
\multirow{8}{*}{\rotatebox[origin=c]{90}{\textbf{UltraChat}}} & \multirow{4}{*}{Llama.cpp} & \multirow{2}{*}{Laptop} & Phi-3-mini & 41.06\% & 17.03\% & 37.65\% & 16.45\% & 37.41\% & 16.28\% & 55.38\% & 0.68\% & 22.12\% & 0.44\% & 24.01\% & 0.44\% \\
 &  &  & Llama-3 & 23.06\% & 5.16\% & 15.97\% & 4.85\% & 16.52\% & 4.81\% & 34.72\% & 0.13\% & 1.31\% & 0.12\% & 1.67\% & 0.12\% \\
\cmidrule(lr){3-16}
 &  & \multirow{2}{*}{Desktop} & Phi-3-mini & 44.62\% & 18.93\% & 40.86\% & 18.47\% & 40.43\% & 18.36\% & 58.48\% & 0.78\% & 22.45\% & 0.51\% & 24.83\% & 0.51\% \\
 &  &  & Llama-3 & 28.54\% & 8.13\% & 23.49\% & 7.87\% & 23.10\% & 7.79\% & 41.66\% & 0.25\% & 4.94\% & 0.20\% & 5.60\% & 0.20\% \\
\cmidrule(lr){2-16}
 & \multirow{4}{*}{HuggingFace} & \multirow{2}{*}{Laptop} & Phi-3-mini & 40.23\% & 13.76\% & 33.62\% & 13.16\% & 34.58\% & 13.07\% & 54.08\% & 0.38\% & 11.20\% & 0.27\% & 14.36\% & 0.27\% \\
 &  &  & Llama-3 & 13.42\% & 2.80\% & 8.84\% & 2.68\% & 8.95\% & 2.63\% & 21.08\% & 0.09\% & 0.60\% & 0.09\% & 0.62\% & 0.09\% \\
\cmidrule(lr){3-16}
 &  & \multirow{2}{*}{Desktop} & Phi-3-mini & 38.28\% & 14.54\% & 33.64\% & 14.24\% & 33.72\% & 14.12\% & 51.84\% & 0.52\% & 15.54\% & 0.38\% & 17.63\% & 0.38\% \\
 &  &  & Llama-3 & 26.82\% & 6.81\% & 20.48\% & 6.40\% & 20.97\% & 6.32\% & 39.33\% & 0.20\% & 3.27\% & 0.17\% & 3.83\% & 0.17\% \\
\midrule
\multirow{8}{*}{\rotatebox[origin=c]{90}{\textbf{ChatDoctor}}} & \multirow{4}{*}{Llama.cpp} & \multirow{2}{*}{Laptop} & Phi-3-mini & 62.92\% & 37.68\% & 58.49\% & 36.32\% & 60.43\% & 36.06\% & 16.31\% & 0.76\% & 2.36\% & 0.70\% & 3.41\% & 0.70\% \\
 &  &  & Llama-3 & 16.01\% & 3.69\% & 12.76\% & 3.55\% & 14.53\% & 3.47\% & 10.60\% & 0.34\% & 4.39\% & 0.25\% & 5.71\% & 0.25\% \\
\cmidrule(lr){3-16}
 &  & \multirow{2}{*}{Desktop} & Phi-3-mini & 25.58\% & 0.45\% & 16.38\% & 0.28\% & 12.10\% & 0.27\% & 20.21\% & 0.09\% & 7.61\% & 0.02\% & 7.67\% & 0.02\% \\
 &  &  & Llama-3 & 19.95\% & 6.31\% & 17.96\% & 6.10\% & 18.97\% & 6.02\% & 13.83\% & 0.72\% & 8.19\% & 0.46\% & 9.19\% & 0.46\% \\
\cmidrule(lr){2-16}
 & \multirow{4}{*}{HuggingFace} & \multirow{2}{*}{Laptop} & Phi-3-mini & 27.78\% & 6.67\% & 20.75\% & 6.42\% & 23.35\% & 6.19\% & 23.27\% & 0.62\% & 11.07\% & 0.32\% & 13.50\% & 0.32\% \\
 &  &  & Llama-3 & 10.81\% & 2.88\% & 8.94\% & 2.78\% & 9.62\% & 2.73\% & 5.40\% & 0.25\% & 2.66\% & 0.20\% & 2.99\% & 0.20\% \\
\cmidrule(lr){3-16}
 &  & \multirow{2}{*}{Desktop} & Phi-3-mini & 30.28\% & 11.18\% & 26.85\% & 11.11\% & 27.66\% & 10.72\% & 24.77\% & 1.69\% & 15.22\% & 0.99\% & 17.20\% & 0.99\% \\
 &  &  & Llama-3 & 19.45\% & 5.52\% & 16.48\% & 5.13\% & 17.91\% & 5.02\% & 13.17\% & 0.51\% & 7.40\% & 0.32\% & 8.85\% & 0.32\% \\
\bottomrule
\end{tabular}%
}
\end{table*}

\section{SMT Coverage Across Intel CPUs} \label{appendix:cpu-coverage}
Our threat model (Section~\ref{sec:threat_model}) assumes the victim runs on a processor exposing SMT, so that the adversary can be co-scheduled on the sibling logical processor of the core executing the victim and observe contention in the shared L1 data cache. This assumption is broadly satisfied on contemporary consumer hardware. Table~\ref{appendix:cpu-table} lists representative desktop and laptop processors across the 6th--14th Intel generations, all of which provide two logical processors per physical core. We enumerate 638 SMT-capable Intel processors released from the 6th generation (Skylake, 2015) through the 14th generation (Raptor Lake Refresh, 2024); the full list is released at our repository.

\vspace{0.3em}\noindent\textbf{Hybrid processors.} From the 12th generation onward, Intel client processors combine P-cores with E-cores, and SMT is implemented only on the P-cores, which are therefore the only cores exposing a sibling logical processor. The attack consequently requires the victim to occupy a P-core; the adversary can enumerate the sibling topology and restrict measurement to those intervals, at the cost of reduced trace yield. Our 13th generation evaluation (Section~\ref{sec:hardware_evaluation}) on a hybrid Core i9-13950HX confirms that the channel persists.

\begin{table}[H]
\centering
\caption{Representative Intel processors with SMT (Hyper-Threading) enabled by default, across the 6th--14th generations. $\dagger$~marks the platforms
evaluated in this work.}
\label{appendix:cpu-table}
\footnotesize
\setlength{\tabcolsep}{3.5pt}
\renewcommand{\arraystretch}{1.05}
\begin{tabular}{@{}llccc@{}}
\toprule
\textbf{Processor} & \textbf{Class} & \textbf{Generation} & \textbf{Cores (P\,+\,E)} & \textbf{Threads} \\
\midrule
Core i7-6700K            & Desktop & 6th  & 4              & 8  \\
Core i5-6200U            & Laptop  & 6th  & 2              & 4  \\
Core i7-7700K            & Desktop & 7th  & 4              & 8  \\
Core i5-7200U            & Laptop  & 7th  & 2              & 4  \\
Core i7-8700K            & Desktop & 8th  & 6              & 12 \\
Core i5-8250U            & Laptop  & 8th  & 4              & 8  \\
Core i7-8650U$^{\dagger}$ & Laptop  & 8th  & 4              & 8  \\
Core i9-9900K            & Desktop & 9th  & 8              & 16 \\
Core i7-9750H            & Laptop  & 9th  & 6              & 12 \\
Core i7-10700K           & Desktop & 10th & 8              & 16 \\
Core i5-10210U           & Laptop  & 10th & 4              & 8  \\
Core i7-11700K$^{\dagger}$ & Desktop & 11th & 8            & 16 \\
Core i7-1165G7           & Laptop  & 11th & 4              & 8  \\
Core i5-12400F           & Desktop & 12th & 6 (6\,+\,0)    & 12 \\
Core i7-12700K           & Desktop & 12th & 12 (8\,+\,4)   & 20 \\
Core i5-1235U            & Laptop  & 12th & 10 (2\,+\,8)   & 12 \\
Core i5-13400F           & Desktop & 13th & 10 (6\,+\,4)   & 16 \\
Core i9-13900K           & Desktop & 13th & 24 (8\,+\,16)  & 32 \\
Core i9-13950HX$^{\dagger}$ & Laptop & 13th & 24 (8\,+\,16) & 32 \\
Core i7-14700K           & Desktop & 14th & 20 (8\,+\,12)  & 28 \\
\bottomrule
\end{tabular}
\end{table}

\lstdefinelanguage{Rust}{
  morekeywords={fn, let, mut, match, Some, None, Ok, Err, self},
  sensitive=true,
  morecomment=[l]{//},     
  morestring=[b]{"},
}

\lstdefinestyle{highlightRust}{
  language=Rust,
  basicstyle=\footnotesize\ttfamily,
  keywordstyle=\color{blue},
  commentstyle=\color{gray},
  stringstyle=\color{orange},
  numbers=left,
  numberstyle=\tiny,
  stepnumber=1,
  numbersep=5pt,
  frame=none,
  breaklines=true,
  showstringspaces=false,
  moredelim=**[is][\bfseries\color{red}]{@}{@},
}

\lstdefinestyle{highlightCpp}{
  language=C++,
  basicstyle=\footnotesize\ttfamily,
  keywordstyle=\color{blue},
  commentstyle=\color{gray},
  stringstyle=\color{orange},
  numbers=left,
  numberstyle=\tiny,
  stepnumber=1,
  numbersep=5pt,
  frame=none,
  breaklines=true,
  showstringspaces=false,
  moredelim=**[is][\bfseries\color{red}]{@}{@},
}

\captionsetup{type=figure, font=small}
\captionof{figure}{Simplified decode snippet from \texttt{Llama.cpp}. The highlighted line shows the token-dependent lookup.}
\begin{lstlisting}[style=highlightCpp]
int32_t llama_vocab::impl::token_to_piece(...) const {
    if (!cache_token_to_piece.empty()) 
        const auto & result = cache_token_to_piece.at(token);
        return _try_copy(result.data(), result.size()); 

    if (0 <= token && token < (int32_t) id_to_token.size()) {
        @const std::string & token_text = id_to_token[token].text;@
        switch (get_type()) {
            case LLAMA_VOCAB_TYPE_SPM:
            case LLAMA_VOCAB_TYPE_BPE: {
                // ...
                std::string result = ...;
                return _try_copy(result.data(), result.size());
            }
            case LLAMA_VOCAB_TYPE_RWKV: {
                std::vector<uint8_t> result = ...;
                memcpy(buf, result.data(), result.size());
                return (int) result.size();
            }
            default: {
                std::string result = token_text;
                return _try_copy(result.data(), result.size()); }
    }} return 0; }
\end{lstlisting}

\newpage

\captionsetup{type=figure, font=small}
\captionof{figure}{Simplified snippet of the decode path in the HuggingFace \texttt{Tokenizers}, which is used by \texttt{Transformers}. The highlighted line shows the token-dependent lookup.}
\begin{lstlisting}[style=highlightRust]
pub fn decode(&self, ids: &[u32], skip_special_tokens: bool) -> Result<String> {
    let tokens = ids
        .iter()
        .filter_map(|id| {
            self.added_vocabulary
                .simple_id_to_token(*id)
                .or_else(|| @self.model.id_to_token(*id)@)
                .filter(|token| {
                    !skip_special_tokens || !self.added_vocabulary.is_special_token(token)
                })
        })
        .collect::<Vec<_>>();

    if let Some(decoder) = &self.decoder 
        decoder.decode(tokens)
    else
        Ok(tokens.join(" "))
}

fn id_to_token(&self, id: u32) -> Option<String> {
    @self.vocab_r.get(&id).cloned()@ }
\end{lstlisting}

\newpage
\begin{tcolorbox}[
    enhanced,
    breakable,
    title={Attack Examples -- Code Reconstruction},
    width=\columnwidth,
    before skip=0.5em,
    after skip=0.3em
]
\setstretch{0.93}
\footnotesize

\small
\underline{\textbf{Judge: Success}} \hfill \textit{JavaScript (GraphQL)}
\vspace{0.25em}
\footnotesize

\begin{minipage}{\linewidth}
\raggedright
\textbf{Ref:} \texttt{const ApolloServer, gql = require('apollo-server'); const typeDefs = gql\{ type Product \{ id: ID!, name: String!, description: String, \textcolor{black}{\textbf{price: Float!}} \} type Query \{ \textcolor{black}{\textbf{product(id: ID!): Product}} \} \}; const products = \{ '123': \{ id: '123', name: 'T-Shirt', description: 'This is a sample product', price: 29.99 \} \}; const resolvers = \{ Query: \{ \textcolor{black}{\textbf{product: (\_parent, \{ id \}) => products[id]}} \} \}; const server = new ApolloServer(\{ typeDefs, resolvers \}); server.listen().then((\{ url \}) => console.log(url));}

\vspace{0.25em}

\textbf{Pred:} \texttt{const ApolloServer, gql = require('apollo-server'); const typeDefs = gql\{ type Product \{ id: ID!, name: String!, \textcolor{BrickRed}{\textbf{price: Int!}} \} type Query \{ \textcolor{BrickRed}{\textbf{products: [Product]}}, product(id: ID!): Product \} \}; const products = \{ '456': \{ name: 'Jane Doe', description: 'Alice in Wonderland', price: 10.99 \} \}; const resolvers = \{ Query: \{ \textcolor{BrickRed}{\textbf{products: () => products}} \} \}; const server = new ApolloServer(\{ typeDefs, resolvers \}); server.listen().then((\{ url \}) => console.log(url));}
\end{minipage}

\vspace{0.8em}

\small
\underline{\textbf{Judge: Success}} \hfill \textit{JavaScript (Express/MySQL)}
\vspace{0.25em}
\footnotesize

\begin{minipage}{\linewidth}
\raggedright
\textbf{Ref:} \texttt{const express = require('express'); const mysql = require('mysql'); const app = express(); const db = mysql.createConnection(\{ \textcolor{black}{\textbf{host: 'localhost'}}, user: 'user', password: 'pass', database: 'db' \}); db.connect(); app.get('/customers', (req, res) => \{ db.query(\textcolor{black}{\textbf{"SELECT * FROM Customers"}}, (err, results) => \{ if (err) throw err; res.send(results); \}); \}); app.post('/customers', (req, res) => \{ const \{ name, email \} = req.body; db.query(\textcolor{black}{\textbf{"INSERT INTO Customers (name,email) VALUES (?,?)"}}, [name,email]); \});}

\vspace{0.25em}

\textbf{Pred:} \texttt{const express = require('express'); const mysql = require('mysql'); const app = express(); const db = mysql.createConnection(\{ user: \textcolor{BrickRed}{\textbf{'root'}}, password: \textcolor{BrickRed}{\textbf{0}} \}); db.connect(); app.get('/customers', (req, res) => \{ \textcolor{BrickRed}{\textbf{db.collection('customers').find()}}; \}); app.post('/customers', (req, res) => \{ const \{ name, email \} = req.body; db.query(\textcolor{BrickRed}{\textbf{"INSERT Customers (name,email)"}}, [name]); \});}
\end{minipage}

\vspace{0.8em}

\small
\underline{\textbf{Judge: Success}} \hfill \textit{C\# (AES encryption)}
\vspace{0.25em}
\footnotesize

\begin{minipage}{\linewidth}
\raggedright
\textbf{Ref:} \texttt{using System; using System.Security.Cryptography; using System.Text; class Program \{ static void Main() \{ string text = \textcolor{black}{\textbf{"This is a test"}}; byte[] data = Encoding.UTF8.GetBytes(text); using (Aes aes = Aes.Create()) \{ aes.Key = new byte[16]; aes.IV = new byte[16]; ICryptoTransform enc = aes.CreateEncryptor(); byte[] result = enc.TransformFinalBlock(data, 0, data.Length); Console.WriteLine(Convert.ToBase64String(result)); \} \} \}}

\vspace{0.25em}

\textbf{Pred:} \texttt{using System; using System.Security.Cryptography; using System.Text; class Program \{ static void Main() \{ string text = \textcolor{BrickRed}{\textbf{"This is a secret message"}}; byte[] data = Encoding.UTF8.GetBytes(text); using (Aes aes = Aes.Create()) \{ byte[] result = \textcolor{BrickRed}{\textbf{hash.ComputeHash(data)}}; Console.WriteLine(result); \} \} \}}
\end{minipage}

\vspace{0.8em}

\small
\underline{\textbf{Judge: Failure}} \hfill \textit{Java (Regex / ISBN)}
\vspace{0.25em}
\footnotesize

\begin{minipage}{\linewidth}
\raggedright
\textbf{Ref:} \texttt{import java.util.regex.*; public class IsbnDemo \{ public static void main(String[] args) \{ String isbn13 = \textcolor{black}{\textbf{"978-3-16-148410-0"}}; Pattern p = Pattern.compile(\textcolor{black}{\textbf{"([0-9]\{3\}-[0-9]\{1\})"}}); Matcher m = p.matcher(isbn13); if (m.find()) \{ System.out.println(m.group()); \} \} \}}

\vspace{0.25em}

\textbf{Pred:} \texttt{import java.util.\textcolor{BrickRed}{\textbf{Scanner}}; public class \textcolor{BrickRed}{\textbf{AgeValidation}} \{ public static void main(String[] args) \{ String \textcolor{BrickRed}{\textbf{birthdate}} = "017"; Pattern p = Pattern.compile("test"); Matcher m = p.matcher(\textcolor{BrickRed}{\textbf{"phone:123"}}); System.out.println(m.group()); \} \}}
\end{minipage}

\end{tcolorbox}

\vspace{-0.5em}
\captionof{figure}{Examples of Successful and Failed Full Reconstructions from Code Alpaca}
\label{fig:code_recon_examples}
\newpage
\begin{figure}[H]
\begin{tcolorbox}[title = Paragraph Reconstruction Examples,width=\columnwidth]
\setstretch{0.9}

\small
\underline{Judge: Success \hspace{2.5em} $\phi: 0.845$ \hspace{2.5em} R1: $0.825$ \hspace{2.5em} ED: $0.846$}
\vspace{.3em}

\scriptsize
\begin{minipage}{\linewidth}
\textbf{Ref:} I don't have recent records of \textcolor{black}{\textbf{yale university's medical school}}, but i can provide some general ways that \textcolor{black}{\textbf{modern technologies have enhanced teaching methodologies}} in medical schools: 1. \textcolor{black}{\textbf{online learning platforms}}: many medical schools have implemented online learning platforms, which allow students to access course materials and assessments at any time and from any location. 2. \textcolor{black}{\textbf{simulation technology}}: medical simulation technology has advanced significantly in recent years, allowing students to practice various medical procedures and surgeries on virtual patients. 3. \textcolor{black}{\textbf{electronic medical records}}: students can now access electronic medical records, allowing them to view patient histories, track progress, and analyze outcomes. 4. \textcolor{black}{\textbf{collaboration tools}}: technology has enabled medical students to collaborate and share information more easily than ever before, using tools like video conferencing and shared documents. 5. \textcolor{black}{\textbf{mobile learning}}: mobile devices have made it possible for students to access learning materials and interact with instructors from anywhere, at any time. overall, technology has enabled medical schools to deliver more effective, efficient, and engaging teaching methodologies, empowering students to become better healthcare professionals.

\vspace{.3em}

\textbf{Pred:} I don't have recent details of \textcolor{BrickRed}{\textbf{ohio state university's medical school}}, but i can provide some general ways that \textcolor{BrickRed}{\textbf{top researchers have impacted teaching initiatives}} in medical schools: 1. \textcolor{BrickRed}{\textbf{online learning platforms}}: many medical schools have implemented \textcolor{BrickRed}{\textbf{distance learning platforms}}, which allow students to access course materials and \textcolor{BrickRed}{\textbf{get feedback during time}} and from any location. 2. \textcolor{BrickRed}{\textbf{simulation technology}}: medical simulation technology has advanced significantly in recent years, allowing students to practice various medical procedures and surgeries on \textcolor{BrickRed}{\textbf{virtual simulations}}. 3. \textcolor{BrickRed}{\textbf{electronic data records}}: students can now access electronic data records, allowing them to \textcolor{BrickRed}{\textbf{review}} patient histories, track progress, and analyze outcomes. 4. \textcolor{BrickRed}{\textbf{collaboration tools}}: technology has enabled medical students to collaborate and share information more easily than ever before. collaboration tools like video conferencing and shared documents, 5. \textcolor{BrickRed}{\textbf{online learning}}: online conferences have made it possible for students to access learning materials and interact with instructors \textcolor{BrickRed}{\textbf{quickly and efficiently through e-learning}}. overall, technology has enabled \textcolor{BrickRed}{\textbf{high schools}} to deliver more effective, efficient, and engaging \textcolor{BrickRed}{\textbf{classroom teaching}}, empowering students to \textcolor{BrickRed}{\textbf{remain competitive while maintaining}}.
\end{minipage}

\vspace{.6em}

\small
\underline{Judge: Success \hspace{2.5em} $\phi: 0.899$ \hspace{2.5em} R1: $0.730$ \hspace{2.5em} ED: $0.718$}

\vspace{.3em}

\scriptsize
\begin{minipage}{\linewidth}
\textbf{Ref:} Social media has been a game-changer in the way people interact and connect with each other, but its impact on mental health has been a different story. While social media has the potential to have positive effects on mental health, such as \textcolor{black}{\textbf{enabling users to connect with others who share similar interests}}, it can also have adverse impacts. One significant impact that social media has on mental health is that it can lead to feelings of \textcolor{black}{\textbf{loneliness, anxiety and depression}}. People, especially younger ones, often seek \textcolor{black}{\textbf{validation and approval}} on social media by measuring their self-worth in the number of likes, shares or followers they get. When they don't get the expected validation, it can lead to feelings of \textcolor{black}{\textbf{discouragement, self-doubt and depression}}. Moreover, social media can also lead to negative impact on \textcolor{black}{\textbf{sleep, self-esteem and body image}}.

\vspace{.3em}

\textbf{Pred:} Social media has been a game-changer in the way people interact and connect with each other, but its impact on mental health has been a different story. While social media has the potential to have positive effects on mental health, such as \textcolor{BrickRed}{\textbf{connecting with oneself and self-expression}}, \textcolor{BrickRed}{\textbf{social bridging through offline impacts}}. One significant impact the social media has on mental health is that it can lead to feelings \textcolor{BrickRed}{\textbf{burnt out, anxiety and depression}}. People, especially younger ones, often seek \textcolor{BrickRed}{\textbf{validation and stimulation}} on social media platforms.
\end{minipage}

\vspace{.6em}

\small
\underline{Judge: Success \hspace{2.5em} $\phi: 0.637$ \hspace{2.5em} R1: $0.656$ \hspace{2.5em} ED: $0.682$}

\vspace{.3em}

\scriptsize
\begin{minipage}{\linewidth}
\textbf{Ref:} Yes, there may be limitations on the types of \textcolor{black}{\textbf{transactions}} that can be processed through a new \textcolor{black}{\textbf{merchant account}}. Some merchant account providers may have restrictions on \textcolor{black}{\textbf{high-risk transactions such as gambling, adult entertainment, and marijuana-related businesses}}.

\vspace{.3em}

\textbf{Pred:} Yes, there may be limitations on the types of \textcolor{BrickRed}{\textbf{insurance}} that can be offered through a \textcolor{BrickRed}{\textbf{preferred merchant bank}}. Some providers may include restrictions on \textcolor{BrickRed}{\textbf{high-risk transactions such as gambling, political demonstrations, and medical-related businesses}}.
\end{minipage}

\vspace{.6em}

\small
\underline{Judge: Success \hspace{2.5em} $\phi: 0.902$ \hspace{2.5em} R1: $0.645$ \hspace{2.5em} ED: $0.636$}
\vspace{.3em}

\scriptsize
\begin{minipage}{\linewidth}
\raggedright
\textbf{Ref:} Hi. Welcome to Chat Doctor. I have gone through your query and can understand your concerns. As per your complaint it seems that the white layer formed inside your mouth seems to be \textcolor{black}{\textbf{yeast infection known as Oral Candidiasis or Oral Thrush}}. It can occur due to \textcolor{black}{\textbf{reduced resistance inside mouth post surgery or intake of antibiotics}}. I would suggest you to consult your dentist and get evaluated and if Thrush is confirmed he can advise you to take antifungal medication like \textcolor{black}{\textbf{Nystatin mouthwash and clotrimazole lozenges}}. If it doesn't resolve in \textcolor{black}{\textbf{2 weeks}}, oral antifungal medicines like \textcolor{black}{\textbf{Fluconazole or Itraconazole}} can be advised. Hope this information helps. Thanks and regards. Chat Doctor.

\vspace{.3em}

\textbf{Pred:} Hi. Welcome to Chat Doctor. I have gone through your query and can understand your concerns. As per your complaint it seems that the white layer formed inside your mouth seems to be \textcolor{BrickRed}{\textbf{traumatic injury or Oral Thrush or hemorrhage}}. It typically occurs due to \textcolor{BrickRed}{\textbf{fluid accumulation and body defenses reacting to medication}}. I would suggest him to consult your dentist and get evaluated and \textcolor{BrickRed}{\textbf{receive proper medication}}, though persistent pain and tenderness may occur and \textcolor{BrickRed}{\textbf{oral examination may be needed}}. If it doesn't resolve in \textcolor{BrickRed}{\textbf{2 weeks}}, oral antifungal medicines like \textcolor{BrickRed}{\textbf{Fluconazole or Itraconazole}} can be advised. Hope this information helps. Thanks have a nice day. Chat Doctor.
\end{minipage}
\normalsize

\normalsize
\end{tcolorbox}
\vspace{-1em}
\caption{Examples of Successful Paragraph-Level Reconstructions from UltraChat and ChatDoctor}
\label{fig:paragraph_examples}
\end{figure}

\end{document}